\documentclass[preprint]{revtex4}
\pdfoutput=1
\usepackage{color}
\usepackage[applemac]{inputenc}
\usepackage{hyperref}
\usepackage{dsfont} 

\usepackage[firstpage]{draftwatermark}
\usepackage{amsmath,amssymb}

\usepackage{graphicx}
\usepackage{dcolumn}
\usepackage{bm}
\usepackage[applemac]{inputenc}
\usepackage[lofdepth,lotdepth]{subfig}

\def\beq{\begin{equation}}
\def\eeq{\end{equation}}
\def\be{\begin{equation}}
\def\ee{\end{equation}}
\def\bea{\begin{eqnarray}}
\def\eea{\end{eqnarray}}

\def\bb{\bar{b}}
\def\bm{\bar{m}}
\def\bn{\bar{n}}
\def\bc{\bar{c}}
\def\bB{\bar{B}}
\def\bF{\bar{F}}
\def\M{M_{Pl}}

\def\bX{\bar{X}}
\newcommand{\ad}{Affleck-Dine }
\def\tK{\tilde{K}}

\def\hbc{holomorphic bisectional curvature }

\newcommand{\beqn}{\begin{eqnarray}}
\newcommand{\eeqn}{\end{eqnarray}}

\DeclareMathSymbol{\C}{\mathbin}{AMSb}{"43}
\newcommand{\ba}{ \left( \begin{array}{c c} }
\newcommand{\ea}{ \end{array} \right) }

\newcommand{\Ms}{\ensuremath{M^2_{Pl}}}
\newcommand{\mgrs}{\ensuremath{m^2_{3/2}}}

\numberwithin{equation}{section}

\DeclareRobustCommand{\SkipTocEntry}[4]{}

\usepackage{dcolumn}
\usepackage{bm}
\usepackage{hyperref}

\usepackage{amssymb}

\begin{document}


\vspace{5cm}
\title{Towards Constraining  Affleck-Dine Baryogenesis }

\author{David Marsh}
\email{dm444@cornell.edu}

\affiliation{%
Department of Physics, Cornell University,
Ithaca, NY 14853}

\date{\today}

\begin{abstract}
\ad baryogenesis is preconditioned on  a certain structure of non-renormalizable operators in the Kähler potential. In this paper, we geometrically characterize the resulting constraint on the field space Kähler geometry and survey the Lagrangian  for correlated predictions. 
  The \ad field backreacts on the inflaton, and   by mapping the backreacted potential to an inflaton potential with a step, we find that  current CMB data  severely constrain  some  versions of the scenario.  
\end{abstract}



\maketitle
\newpage


\section{\label{sec:intro} Introduction}

While several ostensibly viable mechanisms for generating the observed baryon asymmetry of the universe have been proposed, which of them --- if any --- is actually responsible for the discrepancy between the densities of matter and antimatter  is not known. A beautiful mechanism  of baryogenesis which appears very natural in models with supersymmetry
 was proposed long ago by Affleck and Dine. In the seminal paper \cite{AD}, it was noticed that baryogenesis may proceed through the dynamics of  flat directions, which in a supersymmetric theory, like the MSSM,   generically exist in abundance before supersymmetry is broken.
 These flat directions may lead to several  important cosmological consequences (for reviews see \cite{Dine:2003ax, Enqvist:2003gh}), out of which baryogenesis is perhaps the most spectacular.
 
The \ad scenario provides a robust mechanism for baryogenesis and can easily produce a significant baryon number, large enough to reproduce the observed ratio of baryons over photons,
\be
n_B/n_{\gamma} \simeq 6 \cdot 10^{-10} \, ,
\ee
even in the presence of late-time entropy releases.  Intriguingly, the viability of the mechanism is contingent upon the structure of Planck-suppressed operators in both the Kähler potential and the superpotential, thus  providing a window of sensitivity to high-scale physics like string theory.  For several interesting examples of inflation in string theory where these operators can be computed, \ad baryogenesis is impossible. Therefore, if this mechanism would be observationally confirmed, it  would provide important information guiding string theory constructions of inflation and the standard model.   

 A number of correlated predictions of the  \ad mechanism have been noted, including  the prediction of  so called Q-balls \cite{Kusenko:1997si} and --- as have received much attention recently ---  the ease with which the near equivalence of the abundance of dark matter and baryons can be explained in this framework  \cite{DMgen},  see also  \cite{Dine:2003ax}. Nonetheless, it can been argued that due to its apparent robustness  together with the wide range of possible resulting baryon numbers, the scenario can be quite hard  to falsify, thus making it  less attractive as a physical theory.  

The purpose of this paper is to further explore the correlated predictions on the \ad scenario. In doing so, we give a clear geometric characterization of the conditions under which the mechanism is viable in \S\ref{sec:geom}, and exploit the nontrivial structure of ${\cal N} =1$ supergravity to extract correlated predictions for various couplings in the Lagrangian in \S \ref{sec:corr2}. 

Furthermore, a potentially interesting consequence of the \ad scenario is the `backreaction' of the flat direction  on the inflaton potential, and in \S\ref{sec:feat} we discuss in detail how this can give rise to constraints on the parameters of the model. These constraints necessarily involve multi-field inflation, but here we use a simplified single-field model to give a rough estimate its relevance.   The strongest constraints are obtained from the anomalies in the temperature anisotropies of the cosmic microwave background at $l \sim 20$ and $l \sim 40$, which from the temperature measurements alone are only  marginally significant.  Future observations of the polarization of the cosmic microwave background will decisively determine  the severity of  the constraints that can be imposed on the \ad mechanism from precision cosmology. 

We discuss this consequence for both thermal and non-thermal initial conditions for the flat directions, and in both cases the effect is observationally relevant only  if a flat direction is displaced from the global minimum of the potential at the time when cosmological scales left the horizon. This can be avoided in a number of ways, e.g.~by inflation persisting for much longer than the around sixty e-folds required to solve the horizon problem.

\section{\label{sec:AD} The Affleck-Dine Mechanism}
 
For Affleck-Dine baryogenesis to be successful, it is crucial that  the scalar components $\psi^a$ of one or several chiral superfields (denoted $\Psi^a$)  parametrizing gauge invariant, renormalizably flat directions,  obtain large vacuum expectation values (vevs) with nontrivial phases during inflation.  However, there are a number of effects that can trap the flat direction at the origin in field space, inhibiting the subsequent baryogenesis \cite{DRT}. For instance, thermal effects will aspire to achieve a configuration of thermal equilibrium with vanishing condensate vev through Yukawa and gauge interactions, leading to a potential for any flat directions of the form $V(\psi, \bar \psi) \sim T^2 |\psi|^2$.  Furthermore, soft masses of the order of the electroweak scale  give rise to contributions to the potential of the order $ V(\psi, \bar \psi) \sim m^2_{soft} |\psi|^2$. Even more important are the Hubble induced masses of the form $V(\psi, \bar \psi) \sim H^2 |\psi|^2$, which for the simplest possible Kähler potentials corresponding to flat field space geometry,  rapidly and classically evolve any initially displaced flat direction vev to the origin in field space 
and efficiently prohibit the development of a non-vanishing condensate.

Nevertheless, in the formative paper \cite{DRT}, Dine, Randall and Thomas   demonstrated that  by adjusting the numerical coefficients of certain non-renormalizable operators in the Kähler potential, the sign of the Hubble induced mass-squared can be changed, thus giving rise to a tachyonic contribution to the total mass of the condensate at the origin in field space.  For a non-vanishing initial vev of the flat direction at the beginning of inflation, the thermal interactions freeze  out\footnote{More precisely, the freeze-out of thermal interactions  requires the initial vev of the scalar component of the flat direction to be $ \psi_{in} > T/g$, as discussed in \cite{DRT}.  }, thus removing the thermal contribution to the scalar potential. Moreover, since for most models  ${\cal O} (m^2_{soft}) \approx m^2_{EW} \ll H^2$ during inflation, the soft terms are negligible in comparison to the Hubble induced contribution. The resulting tachyonic mass of the \ad field causes a prompt development of a significant condensate vev. The flat direction 
eventually settles down, stabilized by contributions to the potential arising from 
Planck-suppressed non-renormalizable operators in the superpotential, or, in the absence of a superpotential for the flat direction to all orders, by non-renormalizable operators in the Kähler potential.

To be explicit, a field lifted by operators of dimension $n \geq 4$ in the superpotential, 
\be
W \supset \frac{\lambda}{n} \frac{\Psi^n}{M^{n-3}_{Pl}} \, ,
\ee
 will have a scalar potential in which the dominant contributions during $F$-term inflation are given by\footnote{Here we have made the natural assumption that the cutoff scale in the holomorphic superpotential is given by $M_{Pl}$, since lower scales arising from string compactifications typically depend non-holomorphically on moduli.
 }:
\be
V(\psi, \bar \psi) = - c_I H^2 |\psi|^2 + \left( a \frac{\lambda H \psi^n}{n M^{n-3}_{Pl} } + c.c.\right) + |\lambda|^2 \frac{|\psi|^{2n-2}}{M_{Pl}^{2n-6}} \, , \label{Vph}
\ee
where $H$ is the Hubble constant, and $c_I$ and $a$ are constant.  Apart from neglecting the thermal and the electroweak scale soft contributions to the scalar potential as discussed above, here also contributions arising from a nontrivial Kähler potential have been neglected apart from their influence on $c_I$. We will have more to say about this consistent approximation in \S \ref{sec:corr2}.

For $c_I$ and $\lambda$ of order one, the flat direction becomes stabilized at
\bea
\psi_f &\simeq& 
\M \left[ \frac{c_I}{(n-1) |\lambda|^2}          \right]^{\frac{1}{2(n-2)}} \left(\frac{H}{\M} \right)^{\frac{1}{n-2}} \nonumber \\
&\approx& M_{Pl} \left(\frac{H}{M_{Pl}} \right)^{\frac{1}{n-2}}  \, , \label{phi0}
\eea
while the (order $n$) $A$-term of equation \eqref{Vph} fosters a nontrivial phase upon the condensate. For small $c_I$, quantum fluctuations  will contribute to the formation of the condensate.

 After inflation, the (overdamped) condensate tracks an instantaneous minimum of the potential until the effects of the electroweak scale soft terms become non-negligible. At $H \simeq m_{EW}$ the torque exerted on the condensate by the soft $A$-terms triggers  a spiraling motion of the vev towards the origin in field space. If the condensate is charged under a global $U(1)$ baryon symmetry\footnote{In the MSSM the relevant symmetry is the accidental $B -L$, since baryon number by itself is not invariant under non-perturbative sphaleron processes, which are in equilibrium in the early universe. } acting like $\psi \rightarrow e^{i \alpha} \psi$ on the field, the rotation gives rise  to a non-vanishing global charge density e.g,
\be
q = - i \big( \psi^* \dot{\psi} - \psi \dot{\psi^{*}} \big) \, .
\ee
For small vevs, the electroweak scale $A$-terms are subdominant to the soft masses, and the resulting baryon number is approximately conserved.  In \cite{DRT}, the resulting fraction of baryon number over the number of photons was estimated to 
\be
\frac{n_B}{n_{\gamma}} \sim 10^{-10} \left(\frac{T_{R}}{10^9\ {\rm GeV}} \right) \left( \frac{\M}{m_{soft}}\right)^{\frac{n-4}{n-2}} \, ,
\ee
after the  decay of the condensate through thermal scattering with the cosmological plasma. 
Subsequent late-time entropy releases can be necessary in order to achieve a small enough baryon asymmetry.  The vev of the flat direction has also other important consequences for the thermal history of the universe, as discussed in e.g. \cite{Enqvist:2003gh, Other}, and references therein.

\section{\label{sec:geom} Geometric Condition}
The purpose of this paper is to survey the  correlations in the \ad scenario, and, fortunately, the structure of ${\cal N} =1 $ supergravity is highly nontrivial and the condition $c_I \approx 1$ leads to a small array of correlated predictions for various couplings in the Lagrangian. A number of authors have previously discussed the conditions under which \ad baryogenesis is attainable (see e.g.\ \cite{DRT, Casas:1997uk, Kasuya:2006wf, Dutta:2010sg}). In this section we add to this work by obtaining a very transparent representation of the geometric condition in terms of a sectional curvature on the field space manifold, which incorporates the full inflaton dependence of the flat direction mass.

The scalar potential of ${\cal N} =1$ supergravity is given in terms of an effective Kähler potential $K$ and a superpotential $W$ as,
\bea
V = V_F + V_D &=& e^{K/M_{Pl}^2} \left(K^{A \bB} F_{A} \bF_{\bB} - 3 \frac{|W|^2}{M_{Pl}^2} \right) +
 \frac{1}{2} \sum_i g^2_i D^2_i  \, \ , \label{Vsugra}
\eea
with the $F$-terms $F_A = D_A W = (\partial_{A} + \frac{K_A}{\M^2} )W$, where $A$ runs over all the chiral superfields $\Phi^A$ in the theory.  We have included the $D$-term potential for Abelian gauge groups  in equation \eqref{Vsugra}, with $D_i = \phi^A q^i_A K_A + \xi_i$, where $\phi^A$ is the scalar component of $\Phi^A$, $q^i_A$ is the $U(1)_i$ charge of $\phi^A$ and $\xi^i$ is the corresponding field dependent Fayet-Iliopoulos term.

The mass-term for the flat directions arising from the  potential \eqref{Vsugra} is easily obtained from first principles by following the standard prescription for computing soft masses in supergravity \cite{KL}: The chiral fields can be separated into the set of visible sector fields $C^a$ (including the flat directions) and the hidden sector fields $X^m$ (including the inflaton), where, in the visible sector, the $F$-terms  are assumed to vanish and  the vevs are taken to be small compared to $\M$.  It is then appropriate to  make a partial Taylor expansion of the Kähler potential and superpotential around the origin in  the visible sector field space:   
\bea
K &=& \tilde{K} + \tilde{K}_{a \bb} C^a C^{\bb} + \frac{1}{2} ( \tilde{K}^{(2,0)}_{a b} C^a C^b + c.c.\ ) + \ldots \, , \nonumber \\
W &=& \tilde{W} + \frac{1}{2} \mu_{a b} C^a C^b + \frac{1}{6} \lambda_{a b c} C^a C^b C^c + \ldots \, ,
\eea
where all the expansion coefficients are function of the hidden sector fields, e.g.\ $ \tilde{K} = \tilde{K}(X, \bX)\, , \, \mu_{ab} = \mu(X)_{ab} $. Additional information about the chiral and gauge structure of the theory needs to be supplied separately; for instance in the MSSM, the only non-vanishing constant contribution to the $\mu$-term  in the superpotential allowed by gauge and R-parity invariance is $H_u H_{d}$. 
The same symmetries similarly severely restrict the allowed operators in the Kähler potential. 

We will be particularly interested in the \emph{renormalizably flat directions} of the globally supersymmetric theory, for which no gauge and R-parity invariant operator in the superpotential below order $n \geq 4$ exist. In the MSSM, the numerous gauge invariant monomials corresponding to single flat directions were listed in \cite{Gherghetta:1995dv}, and recently in full detail in \cite{Basboll:2009tz}.

The renormalizable superpotential gives rise to interactions between different independently flat directions, so that the presence of a vev of one field may lift many others.
We will subsequently refer to a set of flat directions that remain renormalizably flat in the presence of arbitrary vevs of the other elements of the set, as a \emph{sector} of flat directions.  The different sectors can be studied by constructing gauge invariant polynomials, as discussed in \cite{OtherFlat, Enqvist:2003pb}, where the $H_u L_i$ directions were explicitly constructed.

Denoting the flat directions in some sector by $\Psi^a$, the superpotential in this sector can be written, 
\be
W = \tilde{W}(X) + \frac{1}{n M^{n-3}_{Pl}} \lambda^{(n)}_{a_1 \ldots a_n}(X)\  \Psi^{a_1} \ldots \Psi^{a_n} + \ldots   \label{Wflat}
\ee 
The leading order terms in the resulting Lagrangian are easily obtained \cite{KL}:  under these assumptions the  holomorphic bilinears, holomorphic trilinears  and fermion masses all vanish for vanishing flat direction vevs (i.e.\ $B_{ab}  = 0$,  $A_{a b c} = 0$ and $m_{\psi} =0$ at $\psi^a = 0$), and  the mass matrix for the scalars with kinetic terms given by $\tilde K_{a \bb}$ is given by, 
\bea
\partial_{a } \bar{\partial}_{\bb}V_F &=&
\left(e^{\tilde{K}/\M^2} \frac{ F^{\bn} \bF_{\bn}}{\M^2}- 2 m^2_{3/2}    \right) \tilde{K}_{a \bb}
- e^{\tilde K/\M^2} F^{\bm} \bF^{n} R_{n \bm a \bb}   \nonumber \\
&=&\left(3H^2-\frac{ V_D}{\M^2}+m^2_{3/2} \right) \tilde{K}_{a \bb}   - e^{\tilde K/\M^2} F^{\bm} \bF^{n} R_{n \bm a \bb}  \, , \nonumber \\     \label{ms} 
\eea
where we have used $V_F + V_D = V \simeq 3 H^2 \M^2$, and  introduced the gravitino mass $m^2_{3/2} = e^{\tK/\M^2} |\frac{\tilde W}{\M^2}|^2$, and the field space curvature $R^{\bm}_{\ \bn a \bb} = \partial_{a} \tilde \Gamma^{\bm}_{\bn \bb}$. Furthermore, here $\tilde{K}^{m \bn} \bF_{\bn} = \bF^{m}$. For a non-vanishing $U(1)$  $D$-term potential, this contribution to the scalar mass matrix should be complemented with the contribution from $V_D$, as discussed in e.g.\ \cite{Casas:1997uk, Dudas:2005pr}.  

The $n$-th order $A$-term of \eqref{Vph} can similarly be worked out from $V_F$, and includes contributions from the superpotential as well as from the  $n$-th order terms in the partial Taylor expansion of the Kähler potential, i.e.\ $\tilde{K}^{(n, 0)}(X, \bar X)\psi^n$. We will have more to say about this coupling in \S\ref{sec:corr2}. 

\subsection{Holomorphic Bisectional Curvature} \label{sec:hbc}

 Specializing to the particularly illuminating case of a single hidden sector field, $\phi$, supporting the energy density of the early universe and driving inflation  (hence, the inflaton) and a single flat direction, $\psi$, while momentarily specializing to the case $V_D = 0$,  we find that the canonically normalized mass for the flat direction at the origin in field space is given by,
 \bea
 m^2_{\psi \bar{\psi}} &=& 3H^2  \left(  \left[1 + \frac{1}{3} \left(\frac{m_{3/2}}{ H} \right)^{2 } \right] 
-  \left[1 +  \left(\frac{m_{3/2}}{ H} \right)^{2 }  \right] \M^2 \tilde{K}^{\phi\bar{\phi}} \tilde{K}^{\psi \bar{\psi}} R_{\phi \bar{\phi} \psi \bar{\psi}} \right) \,  . \  \  \  \    
 \eea
The gravitino mass, $m_{3/2}$, may during inflation take on values as large as $H$ without fine-tuning, and is not necessarily related to the late time gravitino mass. The expression,
\be
B[\phi, \psi]  = - \M^2\tilde{K}^{\phi\bar \phi} \tilde{K}^{\psi \bar{\psi} } R_{\phi \bar \phi \psi \bar{\psi} }   \, ,
\ee
is the (dimensionless) \emph{holomorphic bisectional curvature} between the holomorphic curves --- or, equivalently, between the real planes invariant under the complex structure rotations ---  defined by $\phi$ and $\psi$ respectively, evaluated at $\psi=0$. The holomorphic bisectional curvature, first introduced in  \cite{Goldberg}, is one of the most natural concepts of curvature on a Kähler manifold and has recently proven to be a very useful concept in relating local quantities in complex geometry to global theorems, see e.g.\ \cite{Seshadri}. 

Of direct relevance for \ad baryogenesis is that  the the parameter $c_I$ in equation \eqref{Vph} is given by
\be  \label{cI}
c_I = - 3\left(1+B[\phi, \psi] +  \left(\frac{m_{3/2}}{H}\right)^2 \left(\frac{1}{3} + B[\phi, \psi] \right)  \right) \,  . 
\ee
Since during inflation, $H$ necessarily remains approximately constant and so does $m_{3/2}$ for most reasonable models, the functional dependence of $c_I$ on the inflaton vev $\phi$ is determined solely by $B[\phi, \psi] $.

Thus for $F$-term inflation,  tachyonic masses at the origin in field space require a negative holomorphic bisectional curvature. In particular, in the interesting case when $m_{3/2} \simeq H$  during inflation, \ad baryogenesis is conditioned on   $B[\phi, \psi] \lesssim -\frac{2}{3}$, while for  $m_{3/2}  \ll H$ during inflation, the condition sharpens slightly to $B(\phi, \psi ) < - 1$.

 To illustrate the utility of equation \eqref{cI}, let us consider the canonical example of \ad baryogenesis in $F$-term inflation with $\mgrs \simeq H^2$, previously discussed in e.g.\ \cite{DRT, Casas:1997uk}, in which the transition of the condensate is triggered by a non-renormalizable Kähler potential of the form
 \be
 K = |\phi|^2 + |\psi|^2 + \frac{\beta}{\M^2} |\phi|^2 |\psi|^2 \, .
 \ee
Since the Hubble parameter is approximately constant during inflation, the nontrivial inflaton dependence of $c_I$ comes entirely from $B[\phi, \psi]$, which in this case is given by, 
\be
B[\phi, \psi] = - \left( \frac{\beta}{(1 + \beta |\tilde{\phi}|^2 )^2} \right) \, ,
\ee
where  $\tilde{\phi} = \frac{\phi}{\M}$. Clearly, the condition $B( \phi, \psi) |_{\phi = 0} < - \frac{2}{3}$ translates into $\beta > \frac{2}{3}$, and
equation \eqref{cI}  becomes,
\bea
c_{I} &=&  - 4 \left( 1 +  \frac{3 }{2} B[\phi, \psi]  \right) = - 4 \left( 1 -  \frac{3 \beta}{2 (1 + \beta |\tilde{\phi}|^2 )^2} \right) \, . \label{cIexpl}
\eea
This illustrates a point that we will return to in \S\ref{sec:corr}: $c_I$ is already classically a nontrivial function of the inflaton, and keeping this function sufficiently negative for the condensate to remain displaced during inflation and until $H^2 \approx m^2_{soft}$ results in a nontrivial condition on the Kähler potential.  From the simple example \eqref{cIexpl} it follows  that  for a field excursion of the inflaton larger than $\tilde{\phi} = \sqrt{3/8}$, there is no value of the constant  $\beta$ that can keep the condensate from transitioning back to the origin \cite{Casas:1997uk}. A reincarnation of this point will become particularly important for the discussion of large-field inflation in \S\ref{sec:Large}.

Let us comment on the generalization of the above considerations to multiple  flat directions: clearly  the negative contribution to the masses comes from the Riemann curvature $R_{a \bb \phi\bar{\phi}}$, which certainly is not necessarily proportional to the metric on the moduli space $K_{a \bb}$.  It follows that the matrix\,   $(c_I)_{a \bb}$ is in general not universal  or even diagonal.

Let us end this section by commenting on the case when a non-vanishing  $D$-term potential is included, and the total mass for the flat direction is  given by,
\bea
  m^2_{\psi \bar{\psi}} &=& m_{3/2}^2\big( 1 + 3 B[\phi, \psi] \big)  +  3 H^2 \Big(1 - \frac{V_D}{V} \Big) \big(1 + B[\phi, \psi] \big)  +  \tilde{K}^{\psi \bar{\psi}} \partial^2_{\psi \bar{\psi}} V_D\, . \label{mD}
 \eea
If the \ad field is charged under the anomalous $U(1)$ in $V_D$, the last term of \eqref{mD} can contribute with a mass-squared of order $\frac{V_D}{\Ms}$ of either sign. 

If the \ad field is uncharged under the anomalous $U(1)$, then the contribution to the mass matrix from the $D$-term potential, i.e.\ $\tilde{K}^{\psi \bar{\psi}} \partial^2_{\psi \bar{\psi}} V_D$, will arise only at loop-order. The relative magnitude of the  $D$-term potential and the total potential  affects the $F$-term contribution to the flat direction mass, which for a single flat direction can be written, 
\be
c^{(F)}_I  = - 3\left( \frac{4}{3} - \frac{V_D}{V} + \left(2 - \frac{V_D}{V}\right) B[\phi, \psi] \right) \, . \label{cfD}
\ee
In this case, still assuming $\mgrs \simeq H^2$ and now specializing to the case $\partial^2_{\psi \bar \psi} V_D \ll H^2$, the origin in field space become unstable for the flat direction for 
\be
B[\phi, \psi] < - \left(\frac{4 - 3 (\frac{V_D}{V})}{6 - 3 (\frac{V_D}{V})} \right) \, ,
\ee
which for $\frac{V_D}{V} < \frac{4}{3}$ bounds $B[\phi, \psi]$ from above by some negative number. In the window $\frac{V_D}{V} \in (\frac{4}{3}, 2)$, \ad baryogenesis may proceed with a positive holomorphic bisectional curvature. Examples of the \ad mechanism in this range can be constructed --- at least in field theory --- by lifting an AdS minimum of $V_F$ by a $D$-term potential, to the positive energy density of the inflationary epoch. Finally, we note that for a vanishing expectation value of the $F$-term potential, i.e.\ $V_F = 0$, the mass of the flat direction is $(1 + 3 B[\phi, \psi] )m_{3/2}^2 + \tilde{K}^{\psi \bar{\psi}} \partial^2_{\psi \bar{\psi}} V_D$, where the first term arises from nontrivial derivatives on the F-term potential that only vanish in the $W \rightarrow 0$ limit.

\subsection{String Theory Examples} \label{sec:string}
We have shown that the \hbc determines the mass of the flat direction in supergravity and that a successful \ad baryogenesis places certain upper bounds on $B[\phi, \psi]$. Since the connection between the  inflaton and the standard model provided by $B[\phi, \psi]$ arises from Planck-suppressed operators in the K\"ahler potential, the mechanism provides a window of ultraviolet sensitivity to string theory, in which such operators, at least in principle, can be computed. In this section we will illustrate how this can be done through  examples of inflation in string compactifications, demonstrating that $B[\phi, \psi]$ is not an arbitrary  function of the inflaton vev in some well-defined string constructions, and that \ad baryogenesis in fact can be shown to be impossible in broad classes of scenarios.  Consequently, if it can be established that the \ad mechanism is indeed responsible for the observed baryon asymmetry of the universe, this would serve as a nontrivial selection criterion for string theory realizations of inflation and the standard model.

First, let us review the feasibility of \ad baryogenesis in the volume modulus inflation of \cite{Conlon:2008cj}, in the Large Volume Scenario, \cite{Balasubramanian:2005zx}, as previously discussed in \cite{Dutta:2010sg}. In this model, the large volume modulus with scalar component $\tau_b$,  is displaced far from its final, approximately Minkowski, vacuum, and during inflation $\mgrs \gg m^2_{EW}$. Assuming that the visible sector is localized in the internal dimensions, and that the visible sector fields have a diagonal Kähler metric, $\tilde{K}_{a \bb} = \tilde{K}^{(a)} \delta_{a \bb}$, it can be argued that the physical Yukawa couplings should be independent of the overall volume \cite{Conlon:2006tj}.  Since the holomorphic Yukawa couplings are independent of the Kähler moduli to all orders in perturbation theory, it follows from the supergravity formula for the physical Yukawa couplings, 
\be
Y^{\rm Phys.}_{a b c} = e^{\tilde{K}/ 2 \M^2} \frac{Y^{\rm Hol.}_{a b c} }{\sqrt{\tilde{K}^{(a)} \tilde{K}^{(b)} \tilde{K}^{(c)}   }} \, ,
\ee
that the overall volume moduli dependence of the kinetic terms of the visible sector fields is related to $\tilde{K}$ by
\be
\tilde{K}^{(a)} = e^{\tilde K/3 \M^2} \tilde{\kappa}^{(a)}  \, , \label{KLVS}
\ee
where $\kappa^{(a) }$ is independent of the overall volume modulus. This determines the coupling between the inflationary and the visible sector to be,
\be
B[\phi= \tau_b, \psi] = - \frac{1}{3} \, , \label{BLVS}
\ee 
which is identical to the \hbc for no-scale Kähler potentials. It follows  from equation \eqref{cI} that for $F$-term inflation, the contribution proportional to $\mgrs$ famously drops out, while the Hubble induced mass gives,
\be
c_I = -4 -6 B[\phi, \psi] = -2 \, .
\ee
It follows that \ad baryogenesis is not possible for volume modulus inflation, or for any $F$-term inflationary model based on a no-scale Kähler potential \cite{Dutta:2010sg}.

A second, slightly more nontrivial example that to our knowledge previously has not been discussed in the literature is \ad baryogenesis in the context of brane inflation in the KKLT scenario of moduli stabilization \cite{Kachru:2003aw, Kachru:2003sx}. At the level of the four-dimensional effective theory, this system can be modeled by supplementing the supergravity $F$-term potential by an explicitly supersymmetry breaking uplift potential,
\be
V_{\rm tot.} = V_F + V_{up} \, ,
\ee
and  the masses in the visible sector are given by equation \eqref{mD}, after replacing $V_D$ by $V_{up}$. There exists an interesting class of models in which warping has been argued to ensure a sequestered form of the Kähler potential \cite{Randall:1998uk, Kachru:2007xp},
\be
K = - 3 \M^2 \ln \left( -\frac{1}{3}( f_{vis.} + f_{hid.} ) \right) \, , \label{Kseq}
\ee
we again find that, 
\be
B[\phi, \psi] = - \frac{1}{3} \, . \label{Bseq}
\ee
In KKLT, the vev of the $F$-term potential during inflation is approximately given by $V_F \simeq -3 \mgrs \Ms$, and the Hubble constant during inflation can not exceed the gravitino mass  \cite{Kallosh:2004yh}. Writing $\mgrs = (1 + \beta) H^2$, it follows that $\frac{V_{up}}{V} = 2 + \beta$, for some constant $\beta >0$. Using equation \eqref{mD}, we find that 
\be
c_I^{(F)}  = 2 + 2 \beta \, .
\ee 
However, the uplift potential depends on the visible sector fields, and contributes to $c_I$ with \cite{Berg:2010ha},
\be
c_I^{(up)} = - \frac{2}{3} \frac{V_{up}}{\Ms H^2} = -4 - 2\beta \, .
\ee
In conclusion, we find that
\be
c_I = -2 \, ,
\ee
from which it follows that \ad baryogenesis is not possible in brane inflation in KKLT with a sequestered visible sector.  

The \ad mechanism can be embedded in some string models, see e.g.\  \cite{Casas:1997uk}, in which it was noticed that the tree-level K\"ahler potential in orbifold theories gives rise to, in our notation, a constant holomorphic bisectional curvature,
\be
B[T, \psi_{i}] = \frac{n_{i}}{3} \, 
\ee
where $T$ is the overall volume modulus and the candidate inflaton to boot, and $n_{i}$, being an  integer in the interval $[-1,-5]$, is the modular weight of the chiral field $\psi_{i}$. Further examples of the viability of the  \ad mechanism in effective theories coming from string theory can be found in \cite{Dutta:2010sg}. 

The difficulty  in obtaining positive $c_I$ in string theory models of inflation can be traced back to the fact that though the underlying no-scale symmetry is broken in stabilized compactifications, it can still importantly influence the Planck suppressed operators determining the coupling between the visible sector and the inflaton. While in Minkowski space the no-scale structure cause vanishing  scalar masses, in the quasi-de Sitter space relevant for inflation on the other hand, the cancellation is only partial \cite{Gaillard:1995az}, and the resulting mass-squared is always positive.

\section{\label{sec:corr} Correlated Predictions:  \emph{Backreaction on the inflaton}  } \label{sec:feat}
In \S\ref{sec:geom} we have discussed how the necessary condition for \ad baryogenesis can conveniently be written in terms of  $B[\phi, \psi]$ in the single flat direction case, and how this bisectional curvature captures the functional behavior of the mass of the flat direction at the origin in field space as a function of the inflaton vev. In \S\ref{sec:corr2} we will discuss how $B[\phi, \psi]$ appears in other places in the supergravity Lagrangian, and thus gives rise to correlated predictions of the \ad scenario. In this section, however,  we focus on a particular cosmological consequence of the \ad scenario which potentially can severely constrain the mechanism, namely the what can be thought of as the `backreaction' of the transitioning flat direction on the inflaton. Two  questions are especially pertinent for this analysis: 
First, if the \ad field transitions from some initial vev to $\psi_f$ of equation \eqref{phi0} during the circa ten e-folds when the cosmological scales left the horizon, what are the resulting cosmological signatures? Second, to what extent can it be natural to expect a flat direction transitioning during this period? 

The first question closely connects to a large body of work on the cosmological effects of features in the inflaton potential, and, in particular, it partially overlaps with the ``multiple inflation'' scenario of \cite{Adams:1997de}. More generally, a transitioning field gives rise to a short period of multi-field, non-slow roll inflation, and a complete analysis of this period requires an extension of the works of e.g.~\cite{Gordon:2000hv, Peterson:2010np, Senatore:2010wk} to apply also for swiftly turning field trajectories. A full treatment of this multi-field system is a very interesting future direction, here, however, we take a first step towards understanding what cosmological constraints can be placed on these systems by comparing it to known constraints on simpler single-field systems. This simplified analysis can be thought of as modeling the behavior of the \emph{longitudinal} component of the multi-field system, while neglecting the effects of fluctuations transverse to the evolution of the field. We therefore expect that the resulting constraints  from the multiple-field analysis to be even more severe than the corresponding parameter bounds obtained below from the single-field analogy, which motivates the study of these bounds in the simplified system.

  Concretely, by mapping the effects of a transitioning field to a \emph{step} in an effective, single-field inflaton potential, and by using bounds on the height and width of the step from \cite{Hazra} (see also \cite{Adams:2001vc, Peiris:2003ff, Hamann:2007pa, Mortonson:2009qv, Hunt:2004vt, Hunt:2007dn, Joy:2008qd}) extracted from the seven year WMAP data of the CMB temperature anisotropies, we find in \S\ref{sec:num} that the level, $n$, at which the flat direction is lifted in the superpotential and the value of the function $c_{I}$, can both be severely constrained in certain versions of the scenario. 

In \S\ref{sec:IC}, the question of naturalness  is addressed for both small-field and large-field models of inflation, for both thermal and non-thermal initial conditions. For small-field inflation with initially thermally trapped flat directions, we find that observing  traces of a transitioning flat direction through the CMB can be perfectly natural during the first  
$20$ to $30$  e-folds of inflation, and modest fine-tuning can prolong this period substantially. For large-field inflation, transitions might not only be natural, but also abundant during inflation. The near scale-invariance of the CMB on the other hand severely constrains any transitions occurring at random places during  inflation, which forces the Kähler potential  of the large-field model to have a very special form.  

For earlier discussions of the possibility of constraining the initial configuration of the flat directions in the context of \ad baryogenesis using scale invariance, see \cite{Enqvist:1999hv}.

\subsection{Constraining the transitioning field} \label{sec:fit}
While the analysis of the temperature anisotropies of the cosmic microwave background has given substantial evidence that a period of inflation occurred in the early universe, there are still many open questions regarding the exact nature of this period of accelerated expansion. For example, in theoretical models of slow-roll inflation,  certain flatness conditions are enforced on the inflaton potential, \emph{features} in the inflaton potential however,  can be admissible or even favored by the current WMAP data \cite{Peiris:2003ff, Hazra}. In fact, recent analysis of the temperature anisotropies measured by WMAP \cite{WMAP}, QUaD \cite{QUaD} and ACBAR \cite{ACBAR} find an improved fit for inflaton potentials with a small step located at  a specific location of the inflaton potential \cite{Hazra}, consistent with earlier analysis \cite{Peiris:2003ff, Hamann:2007pa, Joy:2008qd}. The size of the step is constrained from data to be no larger than around $.1\%$ in large classes of inflationary models. While these results are intriguing, future observations of the $E$-mode polarization of the CMB spectrum as well as improved bounds on the bispectrum will be able to determine the cosmological significance of these features \cite{Mortonson:2009qv, Hamann:2009bz, Chen:2006xjb, Hotchkiss:2009pj}. 
 
Historically, transitioning flat directions have served as one of the main motivations for the study of localized features in the inflaton potential. Here, we discuss the effects of transitioning flat directions in the context of \ad baryogenesis by mapping the system onto a simplified single-field model with a step in the potential, for which the cosmological constraints  from the CMB spectrum are known. The subset of models of \ad baryogenesis that can be constrained this way partially overlap with the scenario of multiple inflation of  \cite{Adams:1997de}, in which an initially thermally trapped supersymmetric flat direction transitions to a significant vev, and while doing so `backreacts' on the inflaton potential. However, while \cite{Adams:1997de} and the subsequent works \cite{Hunt:2004vt, Hunt:2007dn} focus on a specific small-field inflationary model with a rather small Hubble constant ($H \simeq 10^{-8} \M$), it is certainly interesting to generalize these considerations to broader classes of inflationary models as well as considering more general initial conditions, as we will do in \S\ref{sec:IC}.

To motivate the mapping of the multiple-field system to a single-field system with a localized step, 
 we consider the \ad potential \eqref{Vph} supplemented with an inflaton potential $V_0(\phi)$, supporting small-field slow-roll inflation,
\bea
V(\phi, \psi) &=& V_0(\phi) -c_I(\phi) H_I^2 \psi^2 + |\lambda|^2 \frac{\psi^{2n-2}}{\M^{2n-6}} \, ,
\eea
where we have neglected the order-$n$ $A$-term as well as the phase of the condensate.
The equations of motion of the system are given by, 
\bea
\left\{
\begin{array}{l}
\ddot{\psi}(t) + 3 H(t) \dot{\psi}(t) + \frac{\partial V(\phi, \psi)}{\partial \psi} = 0 \,  \\
\ddot{\phi}(t) + 3 H(t) \dot{\phi}(t) + \frac{\partial V(\phi, \psi)}{\partial \phi} = 0 \,   \\
H(t)^2 = \frac{1}{3 \Ms} \left( \frac{1}{2} \dot{\phi}^2 + \frac{1}{2} \dot{\psi}^2 + V(\phi, \psi) \right) \, .
\end{array}
\right. 
 \label{Syst}
\eea 
While a transitioning flat direction 
affects the cosmological perturbations for a variety of initial conditions, for concreteness let us consider the case when $\psi_i \simeq H$ at some time $t=0$ when the transition begins, and postpone the discussion of initial conditions to \S\ref{sec:IC}. In this case, an analytic solution for $\psi(t)$ is readily obtained at early times, i.e.\ for $\psi \lesssim \psi_f$,
\be
\psi(t) = \psi_i e^{\frac{3 H t}{2} \left( \sqrt{1 + \frac{8}{9} c_I}  -1 \right)   } \, , \label{eq:410}
\ee  
where we have assumed that the transition is prompt so that $H$ and $c_I(\phi)$ are approximately constant during the transition. During the transition, the two-field system $(\phi, \psi)$ evolves from a `ridge' of the potential to settle down in the `valley' at $\psi_f$, much like  a gentle version of the waterfall transition common in models of hybrid inflation (where $c_I \gg 1$, and the transitions terminates the inflationary era). Longitudinal fluctuations of the fields along the instantaneous tangent vector of the field trajectory  give rise to curvature perturbations, while fluctuations orthogonal to the field trajectory result in  entropy perturbations, as discussed in the case of slow-roll inflation in \cite{Gordon:2000hv, Wands:2002bn}. Futhermore, for  $c_I$ of order one and with an unsuppressed dependence on $\phi$, a large-field version of this system has been analyzed in \cite{Abolhasani:2010kn}, where it was noticed that quantum backreaction can typically not be neglected. Extending the two-field analysis to the system of equations  \eqref{Syst} is beyond the scope of this paper. Fortunately however, much can be learned by mapping the two-field system onto a single-field system with a potential with a small step, corresponding to the step induced by the transition of the flat direction along the \emph{longitudinal} motion in the  $(\phi, \psi)$ field space. In this sense, the transitioning flat direction `backreacts' on an effective, longitudinal, single-field inflaton potential to induce a step in it.  

To estimate the steepness of the step in the single-field potential, we note that the duration, $t_{\star}$, of the transition from $\psi_i = H$ to $\psi_f > \psi_i$ of \eqref{phi0}, can be estimated as,
\be
t_{\star} = \frac{2}{3 k H} \ln \left(\frac{\psi_f}{\psi_i} \right) \simeq \frac{1}{H} \left(\frac{n-3}{n-2} \right) \frac{2  }{3 k}\ln\left(\frac{\M}{H}\right) \, , \label{t}
\ee
which provides a lower bound on --- and a good approximation to --- the actual transition time. Here we have abbreviated $k =  \sqrt{1 + \frac{8}{9} c_I}  -1$. Expressed in terms of the longitudinal velocity $|\dot{\phi}_{\parallel}| \equiv \sqrt{\dot{\phi}^2 + \dot{\psi}^2}$, the  slow-roll parameters, 
\bea
\epsilon &=& -\frac{\dot{H}}{H^2} \, , \label{eq:eps} \\
\eta_{\parallel} &=& \frac{\ddot{\phi}_{\parallel}}{H |\dot{\phi}_{\parallel}|} \, , \label{etaP}
\eea
 deviate significantly from their initial, slow-roll, values during the transition. On the other hand, the speed in the $\phi$-direction does not change abruptly during the transition and can be approximated by its initial slow-roll value\footnote{This approximation breaks down if $c_I(\phi)$ contains a linear term in $\phi$ with an order one coefficient, as we discuss below.} $\dot{\phi} \simeq - \sqrt{2 \epsilon_V(\phi_i)} H \M$, where $\epsilon_V = \frac{\Ms}{2} \left( \frac{\partial_{\phi} V}{V} \right)^2$ and $\phi_i$ denotes the inflaton vev just before the transition. The change in the inflaton vev during the transition can then be estimated as,
\bea
\Delta \phi &\simeq& - \sqrt{2 \epsilon_V(\phi_i)}\  \frac{2}{3 k}  \ln \left(\frac{\psi_f}{\psi_i}\right) \ \M  
\simeq - \frac{\sqrt{8/3}}{2.7^2 k} \cdot 10^4 \left( \frac{n-3}{n-2}\right) \ln\left( \frac{\M}{H}\right) \ H \, ,
\eea
assuming that the transition happens reasonably close to the time when cosmological scales left the horizon, so that $\epsilon_V$ is related to the Hubble scale by the COBE normalization,
\be
 \frac{V^{1/4}}{\epsilon^{1/4}} = 2.7 \cdot 10^{-2} \M \, . \label{COBE}
 \ee

The size of the step can readily be obtained by evaluating $\Delta V = V(\phi, \psi_f) - V(\phi, \psi_i)$, for which we find,
\be
\frac{\Delta V}{V_0} \simeq - \left( \frac{n-2}{n-1} \right) \frac{c_I}{3} \left[ \frac{c_I}{(n-1) |\lambda|^2} \right]^{\frac{1}{n-2}} \left(\frac{H}{\M} \right)^{\frac{2}{n-2}} \, .
\ee

Thus, we propose to model  the two-field model as  a single-field model with a step, parametrized as,
\be
V(\varphi) = V_0(\varphi) \left( 1 - c_f \tanh \left(\frac{ \varphi - \varphi_f}{d_f} \right) \right) \, ,
\ee
with size and width approximately given by
\bea
c_f &=&\left( \frac{n-2}{n-1} \right) \frac{c_I}{3} \left[ \frac{c_I}{(n-1) |\lambda|^2} \right]^{\frac{1}{n-2}} \left(\frac{H}{\M} \right)^{\frac{2}{n-2}} \, , \nonumber  \\ \label{cEst}
\\
d_f &\simeq& \frac{\phi_{\parallel}}{4 \M} = \frac{1}{4 \M}  \int_0^{t_{\star}} dt\ \sqrt{\dot \psi^2 + \dot \phi^2}  \label{dEst}
\eea
where $\phi_{\parallel}$ denotes the path length from $(\phi_i, \psi_i)$ to $(\phi_f, \psi_f)$ as measured with the approximately Euclidean field space metric, and is easily evaluated using the approximate solutions for $\psi$ and $\phi$. The single-field model thus corresponds to the longitudinal coordinate along the field trajectory in the $(\phi, \psi)$-plane.

In \S\ref{sec:num}, we perform a more detailed numerical analysis of an example of a transitioning flat direction during inflation, and discuss how the bounds on the parameters $c_f$ and $d_f$ can be interpreted as bounds on $n$ and $c_I$ for a given inflationary model.

\subsubsection{Consequences of an inflaton dependent $c_I$}
While the mapping of the transitioning system to an inflaton potential with a step is well motivated and serves to give a rough idea of what constraints can be imposed on the system, the multi-field system contains a rich variety of physics, that can not all be captured by a single-field inflaton potential with a  step \cite{Shiu:2011qw, Hotchkiss:2009pj}. 

The neglect of multi-field effects is most severe when the fields evolve through a sharp turn in field space, as happens when the  $\psi$ undergo  damped oscillations to settle down at $\psi_f$,
\be
\psi(t) = \psi_f \left(1 + A e^{-\frac{3Ht}{2}} \cos( \omega t + \vartheta) \right) \, , \label{PsiDamped}
\ee
with $\omega^2 = \left(4 c^{(0)}_I (n-2) -\frac{9}{4}\right) H^2$, and for integration constants $A$ and $\vartheta$. Computing the perturbations around this background solutions is an interesting future problem.

Futhermore, the function $c_I(\phi)$ is typically not a constant, as we have argued in \S\ref{sec:hbc}, and can be Taylor expanded in $\tilde \phi = \frac{\phi}{\M}$,
\be
c_I(\phi) = c_I^{(0)} + c_I^{(1)} \tilde \phi + \frac{1}{2} c_I^{(2)} \tilde \phi^{2} + \ldots \label{cexp}
\ee
Once multiplied by the vacuum expectation value of a flat direction, the inflaton dependence in equation \eqref{cexp} leads to corrections to the  inflaton potential in the single-field model, that can not be captured by a step in an otherwise unperturbed inflaton potential. The importance of these corrections can be estimated by considering the effects of a flat direction transitioning during observable inflation, as the unperturbed inflaton potential satisfies the COBE normalization \eqref{COBE}. The unperturbed inflaton potential can be Taylor expanded around $\phi_0$,  such that
\be
V_0(\phi) = 3 H^2 \Ms \left( 1 - a_1\Delta \tilde \phi + \frac{a_2}{2}\left( \Delta \tilde \phi \right)^2 + \ldots  \right) \, , \label{V0}
\ee 
where $\Delta \tilde \phi = \frac{\phi - \phi_0}{\M}$. This expansion is good for all small field models, in which $\Delta \tilde \phi \ll 1$, as well as for some large-field models, e.g. those with  monomial potentials, since in that case the structure of the expansion coefficients $a_i$ ensure that the  true expansion parameter is $\frac{\Delta \phi}{\phi_0}$, which is smaller than unity before the end of inflation. The slow-roll parameters at $\phi_0$ are given by, $\epsilon_V^{(0)} = \frac{1}{2} a_1^2$ and $\eta_V^{(0)} = a_2$. Imposing  the COBE normalization of equation \eqref{COBE} on the inflaton potential \eqref{V0} at $\phi_0$ gives that,
\be
a_1 = \frac{\sqrt{6} \cdot 10^4}{2.7^2} \left(\frac{H}{\M} \right) \, .
\ee 
A flat direction transitioning just as the inflaton passes $\phi_0$ changes the inflaton potential by
\bea
 &\Delta V(\phi)= -\Big(c_I^{(0)}+ c_I^{(1)} \Delta \tilde \phi +& \frac{1}{2} c_I^{(2)} \Delta \tilde \phi^{2} + \ldots \Big) H^2 \left(\frac{H}{\M} \right)^{\frac{2}{n-2}}  .\   \   \   \   \   \   \   \   
  \eea
 The slow-roll parameters change correspodingly: in terms of the Lagrangian cofficients $a_1$ and $a_2$, we find that
 \bea
 \frac{\Delta a_1}{a_1} &=&   \frac{2.7^2}{3 \sqrt{6} } c_I^{(1)} 10^{-4} \left( \frac{\M}{H} \right)^{\frac{n-4}{n-2}}     \, , \label{b1} \\
 \frac{\Delta a_2}{a_2} &=&  \frac{c_I^{(2)}  }{3 \eta_V^{(0)}} \left(\frac{H}{\M} \right)^{\frac{2}{n-2}}   \, . \label{b2}
 \eea
From \eqref{b1}, we find that the condensate vev changes $\epsilon_V$ by an ${\cal O}(1)$ factor if
\be
n \geq 2 \left( \frac{\log\left( |c_I^{(1)}| \left(\frac{\M}{H}\right)^2 \right)-4 }{\log\left( |c^{(1)}_I| \left( \frac{\M}{H} \right) \right)-4} \right) \, , \label{EpsChange}
\ee
while $\eta_V$ is sensitive for transitioning flat directions with an $n$ greater than
\be
n \geq 2 \left(1 + \frac{\log\left( \frac{\M}{H} \right)}{\log \left| \frac{c_I^{(2)}   }{ \eta^{(0)}_V } \right| }\right) \, . \label{EtaChange}
\ee
For example, the large-field model with a quadratic potential $V_0(\phi) = \frac{1}{2} m^2 \phi^2 = \frac{1}{2} m^2 \phi_0^2 \left( 1 - \frac{\Delta \phi}{\phi_0} \right)^2$ and a Hubble scale close to $10^{-4} \M$ is rather insensitive  to corrections in the the tilt ($\epsilon_V$)  but is sensitive  to corrections in the curvature ($\eta$) for condensates transitioning with $n \gtrsim 6$.  Once the condensate has formed, it will in general not be possible to reassemble the inflaton potential to a monomial potential again.

\subsubsection{Backreaction on moduli fields}
 Known ultraviolet-complete models of inflation typically have a  spectrum that contains comparatively light particles with masses around $H$.   
 It is therefore important to understand if  a transitioning \ad field can affect the moduli in a  way  that could possibly give rise to additional observational signatures.
 
 In order to be concrete, we will discuss this question in a particular example based on the ``Kähler moduli inflation'' in the Large Volume Scenario, see \cite{Conlon:2005jm}.
In this case, the mass of the  lightest (and largest) Kähler modulus is given by \cite{Cicoli:2010ha}, 
\be
m^2 \simeq \frac{1}{\ln {\cal V}} H^2 \simeq \frac{1}{15} H^2 \, ,
\ee 
where we have used that the volume is large in string units, ${\cal V} \approx 10^{5}$.

From the special form of the Kähler potential determined by equation \eqref{KLVS},  the  coupling between $\psi^{\dagger} \psi$ and the canonically normalized volume modulus $\Phi = \sqrt{\frac{2}{3}} \ln {\cal V}$, includes the term
\be
V \supset \psi^{\dagger} \psi  \frac{1}{{\cal V}^{2/3} } \frac{V_F}{\Ms} \approx 3 H^2 \M^2 \left( \frac{\psi_f}{\M}  \right)^2 e^{- \sqrt{2/3} \Phi} \, .
 \ee
 If the modulus $\Phi$ is stabilized at the vev $\Phi_0$ with mass $m$ during inflation, the transitioning flat directions induce a shift in the modulus which to leading order in $\delta \Phi/\M$ is given by,
 \be
 \delta \Phi = \sqrt{6}\ \frac{ H^2   }{\tilde{m}^2}  \left(\frac{H}{\M}\right)^{\frac{2}{n-2}} \frac{1}{{\cal V}^{2/3}}  \M  \, , \label{Pshift}
 \ee 
 where we have defined $\tilde{m}^2 = m^2 + 2 H^2 (\frac{H}{\M})^{\frac{2}{n-2}} \frac{1}{{\cal V}^{2/3}}$. 
 Thus for $m \approx H$ during inflation, the shift in the vev of the modulus due to the displaced flat direction is small with respect to $M_{Pl}$, and will not give rise to an observable signature.  In the Large Volume Scenario, the volume suppression of equation  \eqref{Pshift} makes the shift truly negligible. In the broader class of models in which the moduli couples to the flat direction only through the Kähler potential, the suppression in powers of $\big( \frac{H}{\M} \big)$ is  generic, and the resulting shifts in moduli vevs are small.

\subsection{Numerical Analysis} \label{sec:num}
Let us now discuss a specific toy model of  an \ad transition during inflation, in order to demonstrate in detail  how bounds on the size and width of the step in the potential in the single-field model can provide interesting constraints the parameters $n$ and $c_I$ for flat directions transitioning during observable inflation.

Specifically, we consider a two-field potential for the real, scalar, inflaton $\phi$, and the radial component of a flat direction, represented by the real scalar field $\psi$, given by
\bea
V(\phi, \psi) &=&   V_I(\phi) -c^{(0)}_I H_I^2 \psi^2 - c^{(1)}_I \frac{H_I^2}{\M}\ \phi\ \psi^2+ |\lambda|^2 \frac{\psi^{2n-2}}{\M^{2n-6}}  \, ,  \label{Vphipsi}
\eea
where $V_I(\phi)$ is the unperturbed, initial, inflaton potential, $H_I$ is a constant approximately equal to the value of the Hubble parameter during inflation,  and we have included a linear term in the Taylor expansion of $c_I(\phi)$. Considering a small-field model of inflation, for which the inflaton potential, at least locally, can be Taylor expanded as,
\be
V_I(\phi) = V_0 + V_1 \phi+ V_2 \phi^2  \, , \label{V_{I}}
\ee
 we investigate the effects of varying $c_I^{(0)}$ and $n$ for fixed $\lambda$ and for fixed parameters of $V_I(\phi)$. We specialize to the particular case of $H_I \simeq 10^{-6} \M$, and $\eta_V \simeq .67 \cdot 10^{-2}$ initially, while imposing the COBE normalization on the inflaton potential just before the transition. 

For an initial vev of the flat direction, $\psi_i = H_I$,  the initial conditions for the inflaton were chosen so that the system starts out in slow-roll close to the origin in field space. The naturalness of these assumptions will  be discussed in \S\ref{sec:IC}. As $\psi$ condenses, the two-field system slants from the `ridge' in the potential at $\psi=0$, down to the `valley' at $\psi = \psi_f$,  
where it settles down during a short period of  damped oscillations, while simultaneously slowly rolling in the $\phi$-direction.
\pagebreak
\begin{table}[ht]
\centering
\subfloat[Subfigure 1 list of figures text][The evolution of the condensate vev.]{
\includegraphics[width=0.42\textwidth, height=95 pt]{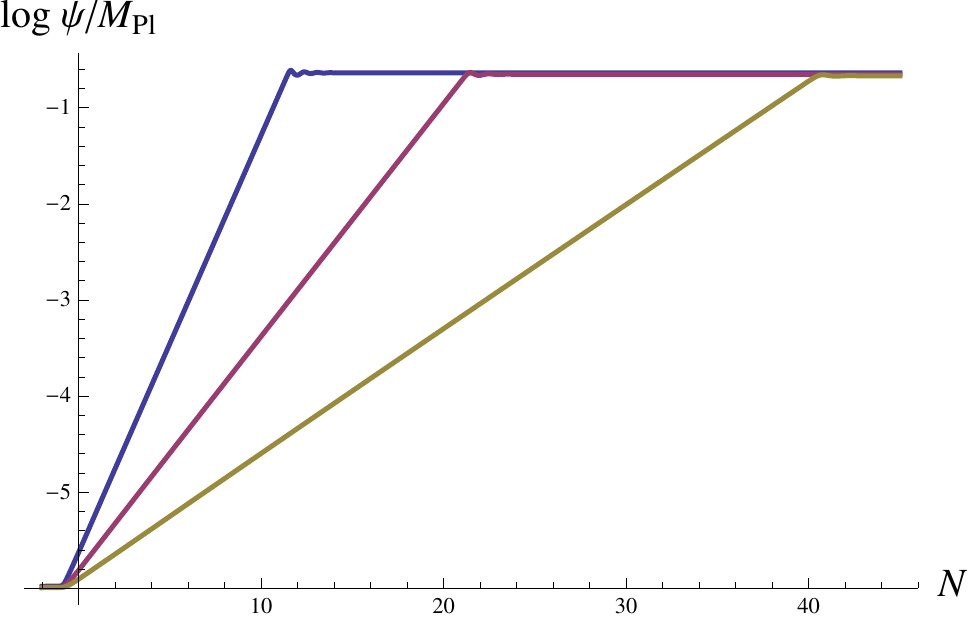}
\label{fig:subfig1}}
\qquad
\subfloat[Subfigure 1 list of figures text][The evolution of the longitudinal speed. ]{
\includegraphics[width=0.42\textwidth, height=95 pt]{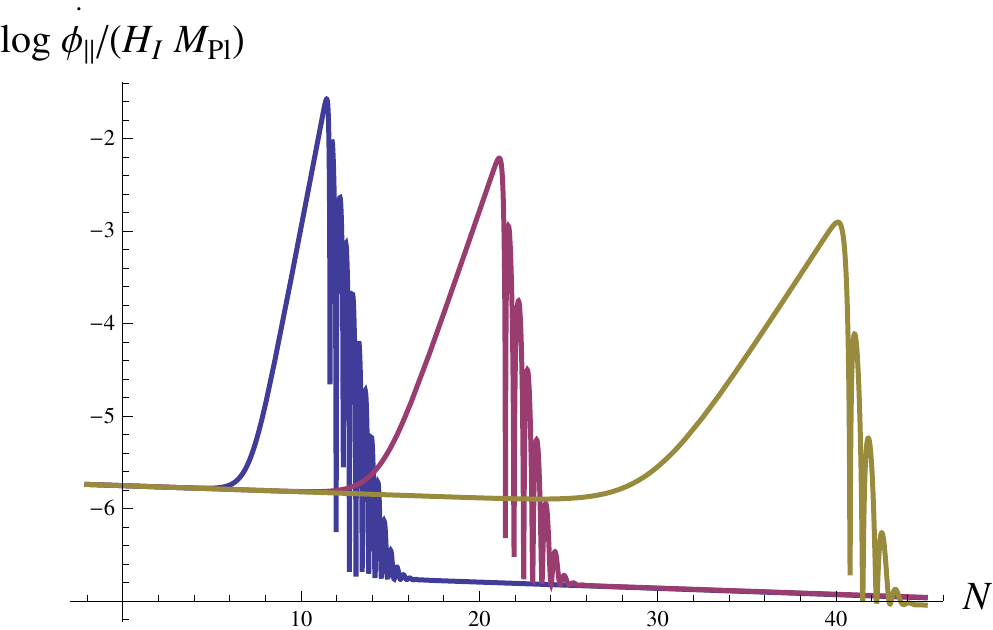}
\label{fig:subfig1}}
\qquad
\subfloat[Subfigure 1 list of figures text][The evolution of the slow-roll parameter $\epsilon_H$ of equation \eqref{eq:eps}.]{
\includegraphics[width=0.42\textwidth, height=95 pt]{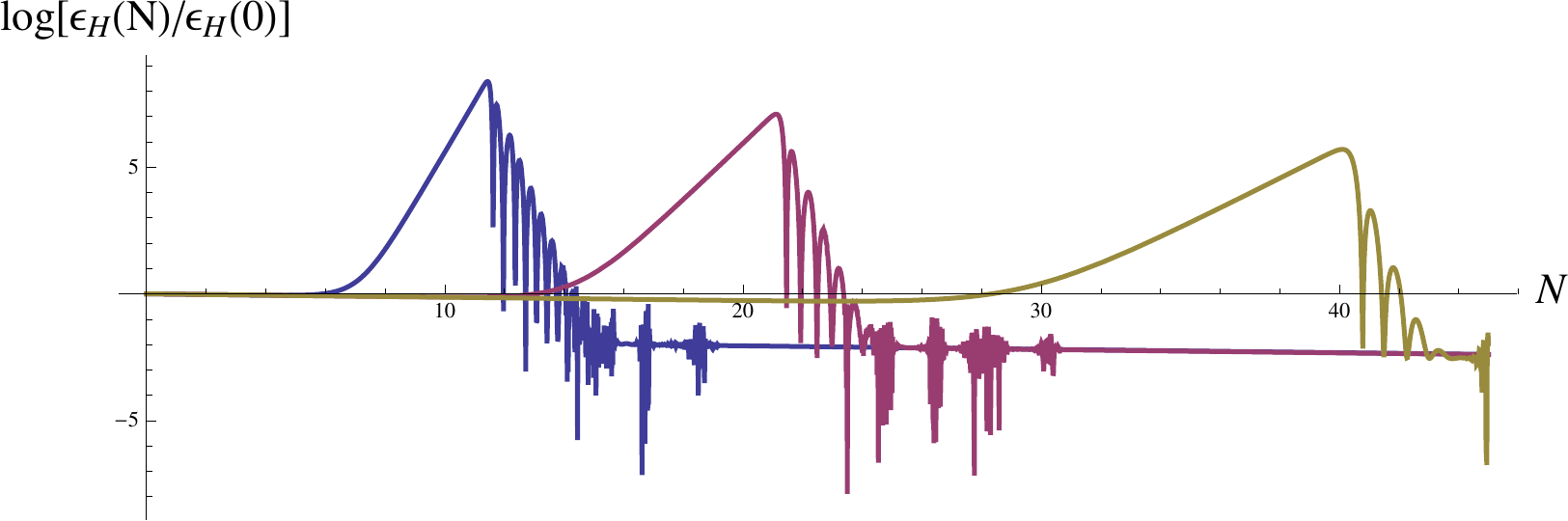}
\label{fig:subfig1}}
\qquad
\subfloat[Subfigure 1 list of figures text][The evolution of the slow-roll parameter $\eta_{\parallel}$ of equation \eqref{etaP}.]{
\includegraphics[width=0.42\textwidth, height=95 pt]{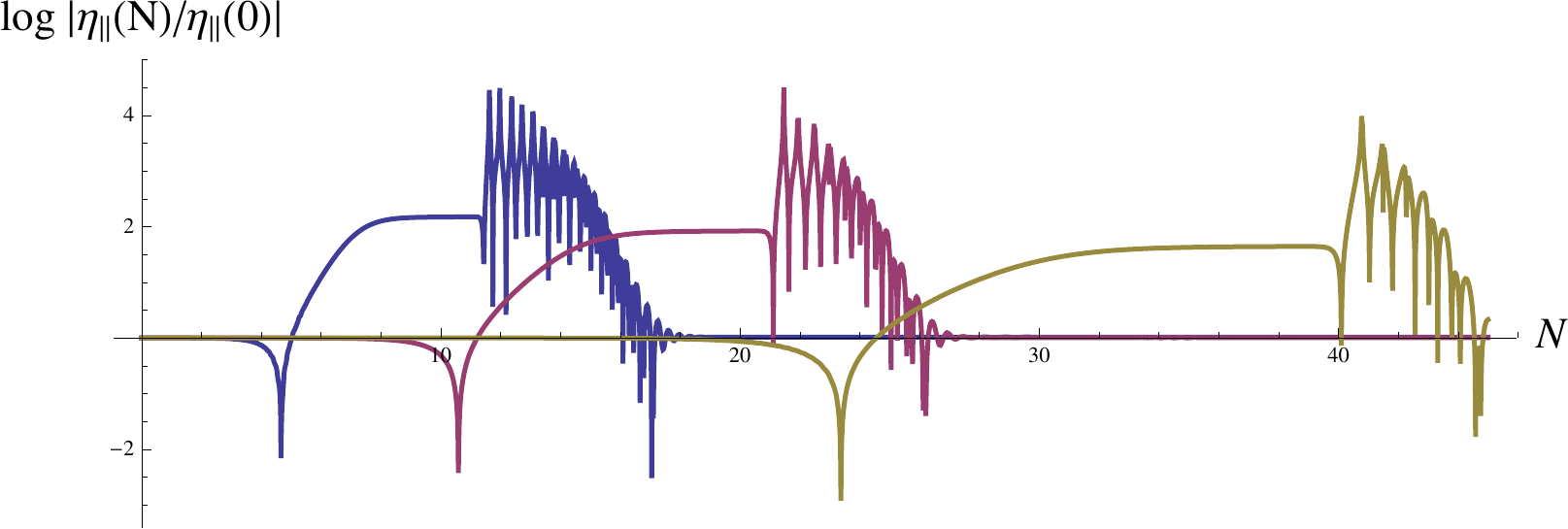}
\label{fig:subfig1}} 
\caption{ The evolution of the condensate vev, the longitudinal speed in field space, and the slow-roll parameters  for $c_I = \frac{1}{2}, 1, 2\   {\rm and}\   n=12$. Please note that the total transition times  are well approximated by equation \eqref{t},  which predicts $H t_{\star} \approx 41, 22,$ and $ 12 $ respectively for each of the parameter values above. Due to the exponential condensation of the flat direction,  the dominant effects of the transition will be localized to a much shorter period of time.
}
\end{table}

As discussed in length in section \ref{sec:fit},  we map the longitudinal projection of this two-field system to a  single-field system  with a  potential with a step, 
\be
V_f(\varphi) =  \left(\tilde{V}_0 + \tilde{V}_1 \varphi+ \tilde{V}_2 \varphi^2 \right) \left(1 - c_f \tanh \left(\frac{\varphi- \varphi_f}{d_f} \right) \right) \, , \label{Vf}
\ee
by determining the parameters $\tilde{V}_0, \tilde{V}_1, \tilde{V}_2, c_f, \varphi_f, d_f$ by numerically fitting $V_f(\varphi)$  to  $V(\phi_{\parallel}\  \big| \phi_{\perp} = 0)$, where $\phi_{\parallel}$ is defined through equation \eqref{dEst}. 

While   $\tilde{V}_0, \tilde{V}_1, \tilde{V}_2$ only differ insignificantly from the inflaton potential parameters  $V_0, V_1, V_2$, the transition inscribes a small step in the overall value of the potential, changing its value by at most a few percent. The system \eqref{Syst} under the influence of \eqref{Vphipsi} can be regarded as a local approximation of the inflaton potential around the location at which the transition occurs, and, as such, it does not describe the full inflationary dynamics until the end of inflation and beyond. In particular, this means that in this simplified model, constraints on the location of the effective step in the inflaton potential (determined by $\varphi_f$) cannot be meaningfully interpreted as a constraint on any model parameters. However, constraints on the width ($d_f$) and the size ($c_f$) of the step do provide constraints on $c_I^{(0)}$ and $n$.

\begin{figure}
\begin{center}
\includegraphics[width=10cm]{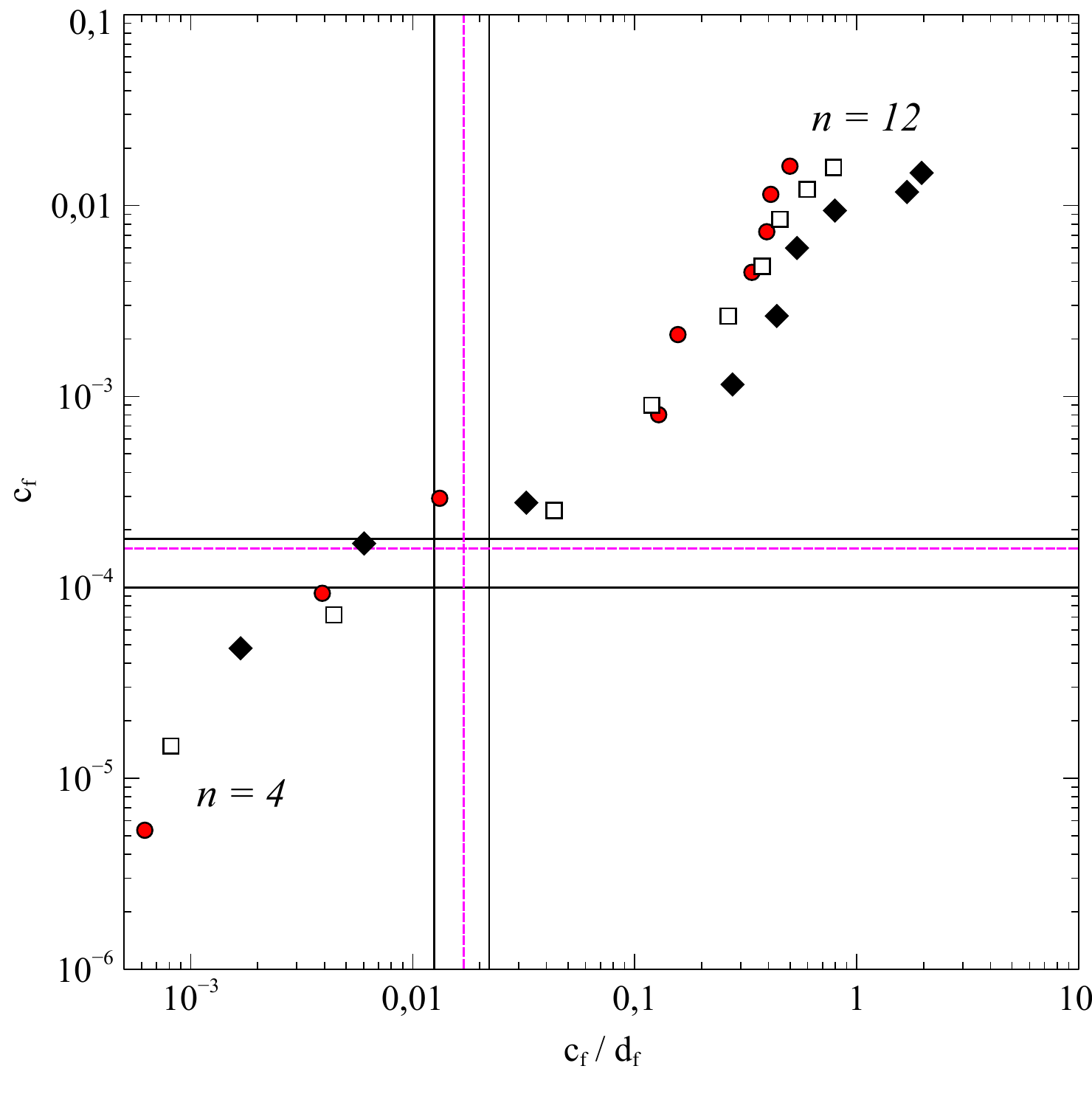}
\caption{
The best-fit value of $c_{f}$ and $\frac{c_{f}}{d_{f}}$ (dashed lines) together with the \emph{constrained} errors for the small-field model of \cite{Hazra}. For $n\in \{4, \ldots, 12\}$, the red circles, white squares and black diamonds correspond to the best-fit values of the parameters for  $c^{(0)}_I  \in \{\frac{1}{2}, 1, 2 \}$, respectively.  In all cases, $c_I^{(1)} =.1$, $\lambda =1$, and the Hubble constant just before the transition is  $H = 10^{-6} \M$, consistent with the COBE normalization of the potential. 
     }
\label{fig_cf}
\end{center}
\end{figure}

In  figure \ref{fig_cf}, we plot the  the best fit values for $c_f$ and $\frac{c_f}{d_f}$ for  $c^{(0)}_I \in \{\frac{1}{2}, 1, 2 \}$ and  $n \in \{4, \ldots 12 \}$,  obtained numerically 
from simulating the system \eqref{Vphipsi} in {\tt Mathematica}, and thereafter fitting $V_f(\varphi)$ to $V(\phi_{\parallel}\  \big| \phi_{\perp} =0)$. 
The dashed lines in figure \ref{fig_cf} correspond to the best-fit-values, i.e.\
 $c_f = 1.6 \cdot 10^{-4}$ and $c_f/d_f = 1.7 \cdot 10^{-2}$, together with the one-sigma constrained errors adapted from one-dimensional sections of the likelihood function of the small-field model in \cite{Hazra}.  Since the constrained errors are given by  sections  of the $68 \%$ confidence curve, they only provide  lower bounds on the errors. In particular, at a larger confidence level, the likelihood function is expected to plateau towards a vanishing step, thus only providing an upper bound on the values $c_f$ and $\frac{c_f}{d_f}$, \cite{Hamann:2007pa}. For $n \lesssim 5$, all points lie in the lower left quadrant, and thus may  not be constrained at a higher confidence level. Figure \ref{fig_cf} illustrates however, that for this subset of \ad models of baryogenesis, bounds on features in the potential can impose severe and interesting constraints on the scenario and may prove useful in singling out  the flat sector responsible for the generation of baryon number.

Let us discuss the limitations of our analysis in detail: While the small-field inflationary model with a step analyzed in \cite{Hazra} using  QUaD, ACBAR  and WMAP five-year as well as seven-year data resulted in an improved fit\footnote{For a discussion of the significance of the improved fit, see \cite{Hamann:2009bz}} with an effective $\chi^2_{eff}$ of about $7$ to $9$, a similar analysis in the two-field case can be expected to differ in many details. In the context of multiple inflation, this question has been addressed in the works \cite{Hunt:2004vt, Hunt:2007dn}, 
where the inclusion of the transitioning field  did not result in an as large improvement of the fit for a model with $H_I \simeq 10^{-8} \M$ and for $n$ taking on half-integer values between $7$ and $9.5$. However, it would be interesting to extend these works to broader classes of inflationary models and to a broad range of transitioning fields. 
 
 The largest limitation to this analysis comes from the particular assumptions made about the initial configuration of the system. We will therefore now turn to the question of naturalness of the chosen initial conditions, as well as the feasibility of extracting correlated predictions from the flat direction backreaction on the inflaton for more general initial configurations. 

\subsection{Initial Conditions \label{sec:IC}}
The transitioning \ad condensate will necessarily  backreact on the inflaton potential; however,  these effects will be observationally irrelevant if the transition ends before the cosmological scales probed by the CMB left the horizon.  For the transition to leave an observable imprint it should terminate between 60 and 50 e-folds before the end of inflation. Clearly, any statement about the naturalness of this happening depends on the physics before observable inflation, out of which very little is known, and what is known is model dependent.  In this section we  discuss different assumptions about the initial configuration of the system, and provide bounds from above on the duration of the period in which transitions are likely. For a determination of the likelihood of a flat direction transitioning during the observable period of inflation, this analysis should be supplemented with an embedding of the inflationary scenario in a UV-complete theory, like string theory. This last task is not addressed  in this paper. 

Independently of baryogenesis,  we find that for large-field inflation, a particular manifestation of the supergravity $\eta$-problem restricts the form of the metric on the visible sector field space to be an approximate solution of the equation,
\be
R_{m \bar n a \bar b} = A\ \tilde K_{m \bar n} \tilde K_{a \bar b} \, ,
\ee
while Affleck-Dine baryogenesis is only viable for certain values of the constant  $A$.

 \subsubsection{Small-field inflation}
 Let us first discuss small-field inflation, such that $\Delta \phi\ll \M$ during inflation \emph{and} all the way until $H \sim m_{EW}$. We will discuss the case when the \ad field starts out in thermal equilibrium with a vanishing vev separately from the case when case when the  vev is large enough for the equilibrating  interactions  to freeze out. The former case is essentially the framework of multiple inflation, discussed in \cite{Adams:1997de}.

  At the onset of inflation, a flat direction can be trapped in a thermal potential  of temperature $T_i$, which can be no larger than $T_i \sim (H_I \M)^{1/2}$ in order for the inflaton to dominate the energy density of the universe. In the inflating background the temperature drops exponentially with the number of e-folds, becoming smaller than the Hubble scale of inflation after $N_T \sim \frac{1}{2}\ln\left(\frac{\M}{H_I} \right) $ e-folds. For Hubble scales between $10^{-8} \M$ and $10^{-4} \M$, this period therefore only lasts for about $5$ to $10$ e-folds. As the temperature drops below the scale $H$, the Hubble-induced terms become important, and the value of the \hbc  between the flat direction and the inflaton--- and thereby $c_I$ --- determines whether the flat direction transitions or not. For $c_I \lesssim -1$, the flat direction remain trapped at the origin in field space even after the thermal effects have ceased to be important. As we have emphasized, $c_I$ is in general a function of the inflaton vev, but for small-field inflation, the leading order contribution in the $\phi/\M$ expansion should well approximate the true value of $c_I$ throughout the inflaton trajectory. 
 
 In case $c_I > 0$, the field will transition as soon as the thermal effects become unimportant. The full transition typically takes several e-folds to complete, and the transition time is logarithmically sensitive to the initial value of the field $\psi_i$. 
 To estimate the initial vev of thermally trapped field, we note that the displacement from the origin will be a result of criticality in the period when the temperature and curvature contribution to the mass approximately cancel. The two-point correlation function in de Sitter space  grows for long wave-length fluctuations \cite{Vilenkin},
 \be
  \langle \psi^2 \rangle  \simeq \frac{H^2}{4 \pi^2} Ht \, .
 \ee
 As an estimate, the period of criticality occurs for between $T^2_i = 2 H^2$ and  $T^2_f \simeq \frac{1}{2} H^2$, which corresponds to period of about a $\ln (2)$ fraction of an e-fold. During this period, the quantum fluctuations of the condensate results in a vev of the condensate  of the order of,
 \be
   \langle \psi^2 \rangle^{1/2} \simeq  \frac{H}{2 \pi} \sqrt{\ln (2)} \approx \frac{1}{10} \  H \, .
 \ee

The analytic solution of $\psi$ for small vevs, equation \eqref{t} then gives  an approximation and  a lower bound for the transition time, 
 \be
 \Delta N \approx \frac{2(n-3)}{3(n-2)} \frac{1}{  \sqrt{1+ \frac{8}{9} c_I  } -1 } \ln \left( \frac{\M}{\psi_i} \right) \, ,
 \ee
  which, as an example, for $\psi_i = H = 10^{-6} \M$, $n=12$, and $c_I = (\frac{1}{2},\ 1,\ 2)$ evaluates to
$\Delta N \approx (12,\  21,\  39)$, in good agreement with numerical simulations, c.f.~Table 1. Furthermore, for $c_I \leq \frac{1}{3}$, the transition period lasts for at least $50$ e-folds for $n\geq 6$, and for more than $60$ e-folds for $n\geq 12$, which demonstrates that transitioning flat directions can also be relevant far past the beginning of the inflationary era. Since the condensation develops exponentially in time, most of the field excursion of the flat direction, and thereby most of the effects of the transition, will be confined to a much shorter period of time of  a few e-folds before and after the end of the transition. As a demonstration of this fact, we note that by the numerical analysis of the preceding section (as is illustrated by the red circles in figure \ref{fig_cf}),  even transitions taking as long as $40$ e-folds in total, give rise to localized features in the effective inflaton potential $V_f(\varphi)$, which are large enough to be easily excluded by CMB data.


For a model like the MSSM with many flat directions, there can be a number of flat directions eligible for transition by all having $c_I> 0$. The field with the maximal $c_I$ will transition first, and in the process, lift all flat sectors it does not belong to through the renormalizable superpotential,  leaving a smaller subset of flat directions still eligible for transition. For a mass estimate of the lifted directions, see \S \ref{sec:int}.   This suggests that $c_I$ may in fact have a typical value that is larger than unity for the flat direction that transitions first. For ${\cal O}(c_I) \simeq 10$, the transition is prompt and over within the first five e-folds after the temperature drops below $H$.  Other flat directions in the same sector may also transition during the same period, and thus may extend the period during which it is reasonable to expect transitioning flat directions.
 
It is interesting to ask whether the transition of a flat direction may ``trigger'' the transitions of other flat directions, which presumably could  extend the total transition period. 
In \S \ref{sec:int}, we discuss the different interactions  in a flat sector, and classify under what conditions such a triggering may occur.  In sum, in the absence of non-holomorphic tri-linears that can induce negative contributions at either quadratic and linear order in the fields, triggering is potentially possible through superpotential couplings between different flat directions. As we discuss in \S\ref{sec:int}, triggering is only possible if these superpotential operators satisfy certain conditions.

 In conclusion, we find that   for  small-field inflation with initially thermally trapped flat directions, a mild fine-tuning of $c_I$ of an order of magnitude is sufficient to extend the transition time to more than $50$ e-folds for most flat directions, and for $c_I = 1$, transitions can be expected during the first $12$ to $22$ e-folds after the thermal potential has subdued --- or equivalently, for the first $20$ to $30$ e-folds of inflation ---  for $4 \leq n \leq 12$.  Since $c_I$ is essentially constant in slow-roll inflation, the number of transitions  is bounded from above by the number of fields in the flat sector that transitions.

Turning to the large class of small-field inflationary model in which the flat directions are \emph{not} thermally trapped at the onset of inflation, we first consider the case where  one or several of the flat directions initially have large vacuum expectation values with $\psi_i > \psi_f$. The relaxation of these fields will again result in a `backreaction' on the inflaton potential. Focusing on the period of slow-roll inflation during which the inflaton $\phi$ is assumed to dominate the energy density of the universe, the initial vevs of the fields are bounded from above by  $\psi_{\rm max} \sim \M \left( \frac{H}{\M} \right)^{\frac{1}{n-1}}$, as discussed in \cite{DRT}. The potential is steeper for fields transitioning from large vevs to $\psi_f$, than for those transitioning from close to the origin in field space, and consequently the period during which it is reasonable to expect  relaxation of flat directions is more limited.  Also in this case, multi-field effects such as ``triggering'' can be very important and may prolong the period during which it is reasonable to expect condensates to develop, but the details are highly model-dependent.

In general, an entire flat sector will transition at the onset of inflation: some to the origin and some to the minimum $\psi_f$ (the magnitude of which may of course differ between the different flat directions, since they can be lifted at different orders and are subject to different coupling constants), depending on the sign of the respective $c_I$'s. If all $c_I \geq 1$, and in the absence of multi-field effects,  these transitions will be over within the  $20$ first e-folds, from which it follows that the backreaction on the inflaton during this period can be substantial. Since multiple fields may transition simultaneously,  the induction  of ``bumps'' in the inflaton  potential is not at all unlikely at the early stages of inflation \cite{Hunt:2007dn}.

 \subsubsection{Large-field inflation \label{sec:Large} }

Let us now discuss the naturalness of observing transitions during large-field inflation, by which we mean inflationary scenarios in which the inflaton field excursion from the onset of inflation and until $H \approx m_{EW}$ is of order $\M$ or larger. Such models suffer more severely from the supergravity $\eta$-problem, and the \ad mechanism analogously requires a very special form of the Kähler potential, as we will now discuss.

An initially thermally trapped field will experience a thermal potential during the first couple of e-folds of inflation after which the exponentially decreasing temperature drops below $H_I$, and the curvature induced masses become important. After this point the fields may transition --- just as in small-field inflation --- if the corresponding \hbc so permits. However, crucially, the holomorphic bisectional curvature (and thereby $c_I$), typically experience an ${\cal O}(1)$ variation over a field excursion of $\M$.  This is exemplified by the simplest possible model of the \ad mechanism in supergravity, cf.\ equation  \eqref{cIexpl}, which successfully  can be embedded in small-field inflation, but not in large-field inflation.

For monomial potentials of the form $m^2 \phi^2$, the field excursion from the last $60$ e-folds and until the inflaton settles down after reheating is a distance of  $15\ \M$ in field space. Thus, during this period it is perfectly natural to expect the value of $c_I$ to change, possibly several times, and for the flat direction to make a number of excursions during inflation. In fact, since in this case transitions every $4$ or so  e-folds may occur, the Affleck-Dine fields will cause significant deviations from a scale invariant primordial spectrum of fluctuations on all scales. Since this is not observed, clearly $c_I$ cannot be a generic function of the inflaton vev and the Kähler potential must be of some restricted form.

Similarly, if the flat directions are not thermally trapped  at the onset of inflation, transitions can be expected at any point during inflation. 

We conclude that in large-field models the transitions of the \ad field are generically ubiquitous, however the non-observance of significant deviations from scale invariance in the primordial power spectrum is suggestive. If the physics of the early universe was indeed  governed by $F$-term large-field inflation, then $B[\phi, \psi]$ should be at least approximately constant, \emph{with or without \ad baryogenesis}. This in turn requires --- at least to a good approximation --- the Riemann tensor to take the form\footnote{Note however that $R_{a \bb c \bar d} $, and $R_{m \bar n l \bar k}$ are not necessarily proportional to the corresponding metric-elements.},
\be
R_{m \bar n a \bar b} = A\   \tilde{K}_{a \bar b} \tilde{K}_{m \bar n} \, , \label{LargeCond}
\ee
evaluated at $\psi^a = 0$, as always. The scalar $A$ is in general an inflaton dependent function, but for $B[\phi, \psi]$ to be slowly varying over Planckian distances, $A$ should similarly be slowly varying and can, at least approximately, be treated  as a constant. In the simplified case of a single visible sector field, $\psi$,  and a single inflaton, $\phi$,   equation \eqref{LargeCond} is the Poisson equation in the flat $\phi$-plane for $f(\phi, \bar \phi) =\ln \tilde{K}_{\psi \bar \psi}$, 
\be
\nabla^{2}_{(\phi)} f(\phi, \bar \phi )  = A \tilde K_{\phi \bar \phi}(\phi, \bar \phi) \, , \label{Aeqn}
\ee
Denoting the Green's function in the domain $D$ of the $\phi$-plane by $G(\phi, \phi')$, we find that
\be
f(\phi, \bar \phi) = A\tilde{K}(\phi, \bar \phi) - A  \oint_{\partial D} ds' \tilde K(\phi', \bar \phi') \partial_n G(\phi, \phi')    \, ,
\ee
where we have assumed that the Green's function vanish on the boundary of $D$. If  $\tilde K$ can be made to vanish on $\partial D$ by a suitable Kähler transformation, the condition of approximately constant masses further simplifies to,   
\be
 \tilde{K}_{\psi \bar \psi}(\phi, \bar \phi) = e^{A\tilde{K}(\phi, \bar \phi)   } \, . \label{Kcond}
\ee
Similar equations can be derived for multiple visible sector fields, if the metric $\tilde K_{a \bb}$ has some special structure.

Motivated by the constant $c_I$ found in the string theory realizations of \S\ref{sec:string}, we can  without any loss of generality write the full Kähler potential of the form $K = - \alpha \ln U$, where $U$ is a real function of the visible and hidden sector fields and $\alpha$ is a constant. In terms of this parametrization of the Kähler potential, the relevant components of the Riemann tensor are 
\be
R_{m \bar n a \bar b} = \frac{1}{\alpha} \tilde{K}_{a \bar b} \tilde{K}_{m \bar n} + V_{a \bar b m \bar n} \, ,
\ee
where 
\be
V_{a \bar b m \bar n} = - \frac{\alpha}{U} \left(U_{a \bar b m \bar n}  - \left(U^{-1} \right)^{\bar c d} U_{a \bar c m} U_{\bar b d \bar n} \right) \, .
\ee
Clearly, the tensor $V_{a \bar b m \bar n}$ has  to be either proportional to the product of the metrics on the moduli space, or vanish in order for the condition \eqref{LargeCond} to hold. For example, in no-scale supergravity and Kähler potentials of the sequestered form, $\alpha = 3$ and $V_{a \bar b m \bar n}$ vanish since the function $U$ is separable, c.f.\ equation \eqref{Kseq}. We discussed in \S\ref{sec:string} how these types of models indeed gives rise to a constant holomorphic bisectional curvature, but do not allow for \ad baryogenesis. Indeed, for $V_{a \bar b m \bar n}$ vanishing, successful \ad baryogenesis bounds the constant $\alpha$ from above,
\be
\alpha < \left\{ \begin{array}{l l}
\frac{3}{2} & {\rm if}\   \mgrs \ll H^2 \, \\
1 & {\rm if}\   \mgrs = H^2 \, . 
\end{array} \right.
\ee
For a single visible sector field and a single inflaton, equation \eqref{Kcond} imply that $U_{\psi \bar \psi}$ is a constant and that the tensor $V_{\psi \bar \psi \phi \bar \phi}$ vanish if $\tilde K$ can be chosen to vanish on $\partial D$.

 \section{\label{sec:corr2} Further Correlated Predictions  } 
The \hbc appears in a number of different places in the supergravity Lagrangian, and the geometric condition of the  \ad scenario discussed in \S\ref{sec:hbc} thus leads to a number of  definite correlated predictions for various couplings.  In this section  we  discuss (classical) wave-function renormalization, fermion terms, and  couplings between multiple flat directions.  

\addtocontents{toc}{\SkipTocEntry}
\subsection{Wave-function normalization}
The function $c_{I}$ of equation \eqref{cI} determines  the canonically normalized mass of a flat direction, and thereby accounts for any changes in the metric on the moduli space $\tilde K_{\psi \bar \psi}$. In the simplified model we used for numerical analysis in \S\ref{sec:num} however, we assumed that the inflaton and flat direction were canonically normalized throughout the relevant part of inflation, despite the fact that the kinetic terms can be expanded as,
\be
\left( 1 - B[\phi, \psi] \left(\frac{\psi}{\M} \right)^{2}  - \frac{1}{4} B[\phi, \phi] \left(\frac{\Delta \phi}{\M}\right)^{2}    \right) \partial_{\mu} \phi \partial^{\mu} \phi^{*} \, ,
\ee
\be
\left( 1 - B[\phi, \psi] \left(\frac{\Delta \phi}{\M}\right)^{2} -  \frac{1}{4} B[\psi, \psi] \left(\frac{\psi}{\M}\right)^{2}  \right) \partial_{\mu} \psi \partial^{\mu} \psi^{*} \, ,
\ee
where the bisectional curvatures are evaluated at some reference point $\phi_{0}$ along the inflationary trajectory at which the fields are canonically normalized, and $\Delta \phi = \phi - \phi_{0}$. This omission is well motivated: for small field inflation $\Delta \phi \ll \Ms$, and the inflaton-dependent correction can be neglected, and while  the transitioning flat direction will induce a change in normalization for both the flat direction itself and for the inflaton, this results in small changes in the parameters of the inflaton potential and negligible changes in $c_{I}$ and $\lambda$ for the flat direction. 
For concreteness, in the model of \S\ref{sec:num} with potential given by equations \eqref{Vphipsi} and \eqref{V_{I}},  the change in canonical normalization of the inflaton due to the transitioning field leads to a redefinition of the parameters of  the inflaton potential of the form
\bea
V_{0} &\rightarrow& V_{0} \, , \\
V_{1} &\rightarrow& \left(1 - B[\phi, \psi] \left(\frac{H}{\M} \right)^{\frac{2}{n-2} }  \right)^{-\frac{1}{2}  }V_{1} \, , \\
V_{2} &\rightarrow& \left(1 - B[\phi, \psi] \left(\frac{H}{\M} \right)^{\frac{2}{n-2} }  \right)^{-1  }V_{2} \, .
\eea
As a result, the inflaton potential for the canonically normalized field will appear slightly tilted after the transition of the flat direction. Nevertheless, the effect of the field-redefinition  is small for most values of $n$, which motivates the omission of this effect in \S\ref{sec:num}.

\addtocontents{toc}{\SkipTocEntry}
\subsection{Fermion couplings}
The Riemann tensor on the field space is well-known to appear in the quartic fermion quartic couplings. The \hbc will thus necessarily gives rise to an additional coupling between the visible sector fermions (denoted $\chi^a$), and fermions in the multiplet of some hidden sector field (denoted $\chi^m$),
\bea
\frac{1}{ \sqrt{- g} }{\cal{L}} &=& - \frac{1}{4}(\frac{1}{\Ms} K_{m \bar{n}} K_{a \bar{b}} + 2 R_{a \bar{b} m \bar{n}}) \chi^m \chi^{\dagger \bar{n}} \chi^a \chi^{\dagger \bar{b}}      \nonumber \\
&=&  - \frac{1}{4 \Ms}\left[1 - 2 B(\phi, \psi) \right] \tilde{K}_{\phi\bar{ \phi}} \tilde{K}_{\psi \bar{\psi}}  \chi^{\phi} \chi^{\dagger \bar{\phi}} \chi^{\psi} \chi^{\dagger \bar{\psi}} 
 \, , \label{fermion} \nonumber \\
\eea
for non-canonically normalized fields. In the last step of equation \eqref{fermion}, we have again specialized to a single flat direction and a single inflaton.  However, even though the \hbc predicted by the \ad mechanism changes the numerical value of this coupling,  it is hardly important for the physics of the early universe.  

For completeness as well as  for the potential embeddings of fermionic preheating in a complete supergravity model,  the masses of the flat direction fermions receive contributions from \emph{both} supersymmetric and supersymmetry breaking sources during inflation, which for the non-canonically normalized field is given by, 
\bea
m_{\psi^a} &=&  e^{K/2 M^2_{Pl}} {\cal D}_{\psi} D_{\psi} W \, .
\eea 
At vanishing  vev of $\psi$,  its fermionic partner is massless, while for a displaced flat direction with vacuum expectation value $\psi_{f}$, the mass is to leading order in $\frac{\psi_{f}}{\M}$ given by,
\bea
m_{\psi } &=& \tilde K _{\psi \bar \psi} \left[ \left(\frac{W}{|W|} m_{3/2} - k_{1} \sqrt{H^{2} + \mgrs} \right) \tilde K _{\psi \bar \psi} (\frac{\psi_f^{*}}{\M})^{2} \right.  \nonumber \\
&+& \left. k_{2} \sqrt{H^{2} +\mgrs} \tilde K _{\psi \bar \psi} |\frac{\psi_f}{\M}|^{2}  \right] \, ,
\eea
where 
\be
k_{1} = \sqrt{3}  \M^{3 }\frac{F_{\phi}}{|F_{\phi}|} \left(  \tilde K ^{\phi \bar \phi}\right)^{1/2} \left(\tilde K^{\psi \bar \psi} \right)^{2}  \partial_{\phi} \tilde K^{(2,2)}_{\psi \bar \psi \psi \bar \psi} \, , 
\ee
and,
\be
 k_{2} =  \sqrt{3} \M^{3} \frac{F_{\phi}}{|F_{\phi}|} \left(  \tilde K ^{\phi \bar \phi}\right)^{1/2} \left(\tilde K^{\psi \bar \psi} \right)^{3} \tilde K^{(2,2)}_{\psi \bar \psi \psi \bar \psi}  \partial_{\phi}  \tilde K_{\psi \bar \psi } \, .
 \ee 
The mass of the canonically normalized fermion is therefore of the order of $ H \left(\frac{H}{\M}\right)^{\frac{2}{n-2}}  $ during inflation.

 After inflation, once  $H \simeq m_{3/2} \simeq m_{EW}$,  the mass of the   flat direction fermion is of the order of $m_{{3/2}}\left(\frac{m_{{3/2}}}{\M} \right)^{\frac{2}{n-2}} \ll m_{{3/2}}$. 

\addtocontents{toc}{\SkipTocEntry}
\subsection{Consequences for multi-field dynamics \label{sec:int}}
Any supersymmetric extension of the Standard Model will  involve many renormalizably flat directions --- in the MSSM they add up to hundreds --- and the \ad mechanism is consequently concerned with a multi-field system. In this section we survey the leading couplings between different flat directions with a particular focus on the interactions  in multi-dimensional flat sectors, as well as the possibility of a developing \ad condensate  ``triggering'' a transition of other flat directions. We discuss how triggering may proceed through superpotential interactions or through  the supergravity induced non-holomorphic tri-linear $C$-term.

To avoid unnecessary cluttering, we will in this section assume that $e^{\tilde K/\Ms} \simeq 1$ and that all fields are canonically normalized. It is not hard to generalize the following equations to the more general case.

\subsubsection{Interactions induced by the renormalizable superpotential}
While any flat direction per definition is $F$- and $D$-flat in global, Minkowski, supersymmetry with respect to the renormalizable superpotential, this does not mean that all flat directions can simultaneously  obtain large vacuum expectation values. The renormalizable superpotential introduces interactions between different flat directions, and the presence of a condensate can significantly restrict the number of dynamically interesting fields.

For example, in the MSSM both the $\Psi_1^2 = H_u H_d$ operator and the $\Psi_2^3 = L_1 L_3 e_1$ operator correspond to supersymmetric flat directions. The lepton Yukawa coupling, 
\be
W \supset \lambda^{ij}_e H_d L_i e_j \, ,
\ee
gives rise to an $F$-term for $L_i^{\alpha}$, (here $\alpha$ is an $SU(2)$ index and $i$ a flavor index), which in global supersymmetry is given by,
\be
F_{L_i^{\alpha}} = H_d^{\alpha} e_i = 0 \, , 
\ee 
where invertibility of the lepton Yukawa matrix has been assumed. Clearly, simultaneous vacuum expectation values of $\Psi_1$ and $\Psi_2$ give rise to a non-vanishing $F$-term.

In general, the renormalizable superpotential contains couplings between different flat directions of the form,
\be
W \supset \lambda \Psi_1 \Psi_2^2 \, ,
\ee
which, in the globally supersymmetric scalar potential,  results in a quartic coupling  of the form
\be
V(\psi_1, \psi_2) \supset 4  |\lambda|^2 |\psi_1|^2 |\psi_2|^2 \, .
\ee

 If either of the fields transitions to a magnitude of $\psi_f$, as in equation \eqref{phi0}, the other will obtain a mass-squared of the order of 
\be
|\lambda|^2 \Ms \left(\frac{H}{\M} \right)^{\frac{2}{n-2}}  = |\lambda|^2 H^2 \left(\frac{\M}{H} \right)^{\frac{n-4}{n-2}}   \, ,
\ee
where $\lambda$ denotes to the relevant Yukawa coupling. For small $n$, and for the smaller of the Yukawa couplings, this contribution of the mass is much smaller than the Hubble constant during inflation, and thus a subleading contribution to the total mass of the flat direction, while for larger $n$ and for Yukawa couplings $\lambda \gtrsim 10^{-3}$, it can lead to induced masses of the order of Hubble or larger. In the former case, we would regard the fields to belong to the same flat sector with respect to the renormalizable superpotential, while in the latter case they would belong to different sectors. To our knowledge, there is no complete classification of simultaneous flat directions, however see \cite{Enqvist:2003pb, OtherFlat} for some interesting special cases.

\subsubsection{Order-$n$ $A$-terms}
The most important contribution from the non-renormalizable superpotential terms of the form \eqref{Wflat} is to generate the stabilizing term proportional to $|\psi|^{2n-2}$ in the scalar potential, as well as the order-$n$ $A$-term, as in \eqref{Vph}. In supergravity, the $A$-term arises from contributions to the scalar potential of the form
\bea
V_F &\supset& \frac{1}{n}  \Big( \tilde{\bar{F}}_{\bar \phi}\tilde{K}^{\phi \bar \phi} (D_{\phi} \lambda) \M^{3}-   3 \lambda \frac{\tilde W^{*}}{|\tilde W|}  m_{{3/2}} \M^{3}        \Big) \   \left( \frac{\psi}{\M}  \right)^n  \, \nonumber  \\
&=&\frac{1}{n} \Big( \sqrt{3 (H^2 + \mgrs)} \frac{\tilde{\bar{F}}_{\phi }}{|\tilde{\bar{F}}_{\phi} |}D_{\phi} \lambda - 3 \lambda \frac{\tilde W^{*}}{|\tilde W|}  m_{{3/2}} \M^{3}       \Big) \   \left( \frac{\psi}{\M}\right)^n \, . \label{Aterm} 
\eea
where we have specialized to a single hidden sector field and neglected higher order terms in Kähler potential of type $\tilde K^{(n, 0)}$ contribute with $\frac{H}{\M}$ suppressed corrections to the coupling \eqref{Aterm}. 

\subsubsection{Triggering}
More interesting are the contributions to the non-renormalizable superpotential that mix different flat directions, as in equation \eqref{Wflat}. Cross-couplings between  flat directions in the same sector can be present at order $k > 3$. For concreteness, suppose that the field $\psi_1$ appears at quadratic order in a contribution to the superpotential at order $k$ which couples it to a second flat direction $\psi_2$, i.e.\
\be
W \supset \frac{\lambda^{(k)}}{k} \M^{3-k} \Psi_1^2 \Psi_2^{k-2} \, . \label{Wmix}
\ee
If $\psi_2$  obtains a vev of the order of $\psi_f$, then the resulting mass-squared for the field $\psi_1$ will obtain a positive definite contribution of the form,
\be
|\lambda^{(k)}|^2 H^2 \left(\frac{2}{k} \right)^2  \left(\frac{H}{\M} \right)^{\frac{2(k-n)}{n-2}} \,
\ee  
The superpotential coupling \eqref{Wmix}  also gives rise to an order-$k$ holomorphic $A$-term in the scalar potential of the type \eqref{Aterm}.  After $\psi_2$ condenses, this coupling will schematically contribute to the $\psi_1$ mass with,
\be
\frac{|\lambda^{(k)}|}{k} H \sqrt{H^2 + \mgrs} \left( \frac{H}{\M }\right)^{\frac{k-n}{n-2} } \cos(\beta) \, , \label{trigg}
\ee
The angle $\beta$ depends on the phases of $\psi_1, \psi_2$ and $D_{\phi} \lambda^{(k)}$, and --- in the absence of other contributions to the mass of  $\psi_1$ --- the cosine dependence of \eqref{trigg} will cause two tachyonic directions to open up in the $\psi_1$ plane, under the condition that
\be
\frac{1}{k} |\lambda^{(k)}| \left( \frac{H}{\M}\right)^{\frac{k-n}{n-2} } \lessapprox 1 \, . \label{TriggCond}
\ee 
For $k < n$, this equation requires fine-tuning of the coefficient $\lambda^{(k)}$ to be satisfied, and triggering due to condensation does not appear to be generic. For $k>n$ on the other hand, equation \eqref{TriggCond} only imposes mild restrictions on the value of $\lambda^{(k)}$ to be satisfied. In this case however, the magnitude of the tachyonic contribution to the mass in equation  \eqref{trigg} is suppressed with respect to $H^2$, and may not render the total mass-squared negative.

\subsubsection{``$C$-terms''}
The phenomenological potential \eqref{Vph} gives a clear view of the dominant effect for \ad baryogenesis involving a single field. For multiple fields however, there are more terms that potentially can produce interesting effects. In particular, the nontrivial Kähler potential can induce  terms at all orders, and thus introduce operators with a much lower dimension than those originating from the  superpotential. For flat directions, the lowest dimensional contributions of this sort  arise at cubic order, in terms of the non-holomorphic tri-linears sometimes called ``$C$-terms'' \cite{Martin:1997ns},
\bea
V_F \supset \frac{1}{2}  c^{(2,1)}_{a b \bc}\  \psi^a \psi^b \psi^{* \bc} + c.c. \, ,
\eea
As we will see,  $C$ typically is of the order $H \frac{H}{\M}$, and the $C$-term gives rise to a mass term of the order of 
\be
H^2 \left(\frac{H}{\M}\right)^{\frac{1}{n-2}} \, ,
\ee 
for a flat direction appearing quadratically in the $C$-term, coupling to another flat direction that is lifted at order $n$ with vev given by \eqref{phi0}. On the other hand,  a field appearing linearly in the $C$-term may experience a \emph{linear} instability  of the order of
\be
H^2 \M  \left(\frac{H}{\M}\right)^{\frac{2}{n-2}} \, ,
\ee
if the other field appearing in the $C$-term have condensed. 

This type of non-holomorphic --- sometimes called ``maybe soft'' --- tri-linears  are severely restricted by gauge invariance and R-parity \cite{Martin:1997ns}. For instance,  in terms of the ordinary MSSM fields, the only operators of this form are \cite{Hall:1990ac} 
\bea
 \tilde{Q} \tilde{u} H^{*}_{d}\, ,\    \tilde{Q} \tilde{d} H^{*}_u \, , \    \tilde{L} \tilde{e} H^{*}_u  \, . \label{Cmssm}
\eea
To assess the importance of these operators, we  first note that all the $C$-terms  involve a Higgs field and thus  couples the only two flat directions including the Higgs fields, $H_u H_d$ and $H_u L_i$,  to operators involving squarks or sleptons. For the $H_u H_d$ direction, $F$-flatness of the renormalizable superpotential requires both $\tilde{Q}_i =0$ and $\tilde{e}_i = 0$, thus enforcing the vanishing of all $C$-terms coupling to the $H_u H_d$-direction. In other words, for the $H_u H_d$ operator, the $C$-terms lift no directions that  are not also lifted by the renormalizable superpotential. 

The $H_u L_i$ direction is simultaneously $F$-flat with some operators of the form $L_i L_j e_k$, and studying  this sector in detail --- including the possibility of a non-vanishing $C$-term --- could be very interesting.

The supergravity induced  $C$-terms can easily be given a geometric interpretation, since when phrased in terms of the covariant fluctuation obtained by the background field method, explained in \cite{Mukhi}, they are given by 
\bea
\hat{c}^{(2,1)}_{a b \bc} &=&   \frac{2}{3}  \left( \nabla_{a} \nabla_{b} \nabla_{\bar c} V_F + \nabla_{a} \nabla_{\bar c} \nabla_{b} V_F + \nabla_{\bar c} \nabla_{a} \nabla_{b} V_F \right) \, = \nonumber \\
&=& 2 \left( \nabla^{3}_{a b \bar c} - \frac{1}{3} R^{B}_{\  a \bar c b} \nabla_{B}\right) V_{F}  \nonumber \\
&=&-2 e^{\tilde K} \Big( F^{\bm} \bF^n \nabla_{a} R_{\bm n \bc  b}   +
  \frac{1}{3} R_{a \bc b \bm}(\tilde{K}^{m \bm} \bF^l {\cal D}_m F_{l}  + F^{\bm} \bar{W}) \Big) \, ,\label{c}
\eea
 where  $ {\cal D}_m F_{l} = \partial_{m} F_{l} + K_m F_{l} - \Gamma^p_{m l} F_{p}$, all in natural units.

Since, by gauge and R-parity invariance, there is no allowed cubic self-interaction from the $C$-terms in the MSSM for any flat direction, we have
\be
\hat{c}^{(2,1)}_{\psi \psi \bar{\psi} } = 0 \, .
\ee
The three terms contributing to $\hat{c}^{(2,1)}_{\psi \psi \bar{\psi} } $ in \eqref{c} are in general independent functions of the inflaton which (depending on the inflationary scenario)  may even be of different orders of magnitude, and we will not consider the case when there are nontrivial cancellations between them. It then follows that each term has to cancel separately so that at vanishing vev of the flat direction,  $\psi = 0$,
\be
R_{\psi \bar \psi \psi \bar{\phi}} = 0 \, , \label{Cond1}
\ee
and  the relevant holomorphic bisectional curvature is covariantly constant along the flat direction,
\be
\nabla_{\psi} B[\phi, \psi] = 0 \, . \label{Cond2}
\ee

\subsubsection{Quartic interactions}

Finally, the quartic interactions couple all flat directions  to each other and generically  gives rise to contributions to the squared masses of a flat direction of the order of  
\be
H^2 \left(\frac{H}{\M}\right)^{\frac{2}{n-2}} \, ,
\ee 
in the background of another condensed flat direction, stabilized at order $n$ in the superpotential. In case a given flat direction is not stabilized by any non-renormalizable operator in the superpotential, the quartic self-interactions  will stabilize the condensate to a vev of the order of $\M$.

The quartic scalar couplings for the covariant fluctuations, $\hat{\psi}^a$,
\be
V_F \supset \frac{1}{4} \lambda_{a \bar b c \bar d}  \hat{\psi}^a  \hat{\psi}^{\bb}  \hat{\psi}^{c} \hat{\psi}^{\bar d} \, ,
\ee
are  given by the symmetrized covariant derivative $\frac{1}{3!} \nabla^4_{( a \bb c \bar d)}V_F$, which can be written in natural units as, 
\bea
 \lambda_{a \bar b c \bar d}  &=& e^{\tilde K} \Big[   \left( F_{m} \bar{F}^{m} - |W|^2 \right) \tilde{K}_{\bb \{a} \tilde{K}_{c \} \bar d}  
 + \bar{F}^m F^{\bar n}\big(   K^{A \bar A}  R_{m \bar n c \bar A}  R_{ A \bb a \bar d } -  K^{A \bar A} R_{m \bar n  A \bar d}  R_{ \bar A  a \bb c }  
 \nonumber   \\
 &+&    K^{A \bar A} R_{m \bar A  \{ a \bar d}  R_{ A \bar n   c \} \bb }     -  K_{\{ a \bb} R_{m \bar n c\} \bar d}
 - K_{\{ a \bar d} R_{m \bar n c\} \bar b} - \nabla^2_{\bb a } R_{m \bar n c \bar d}  \big)  - R_{a \bar b c \bar d} |W|^2  \nonumber  \\
 &-&  W \bar{F}^n \nabla_{a} R_{n \bar d c \bar b}  - \bar{W} F^{\bar n} \nabla_{\bb} R_{\bar n a \bar d  c} \Big]
\eea
where braced indices are symmetrized (i.e, $T_{\{a b\} \ldots } = T_{a b \ldots } + T_{b a \ldots }$), and capital letters run over both the visible and hidden sectors. For a single inflaton and a canonically normalized flat direction, the expression simplifies to
\bea
 \lambda_{\psi \bar \psi \psi \bar \psi} &=& 6 H^{2} \left( \left(1 +  B[\phi ,\psi] \right)^2 +\frac{1}{2} \nabla^{2}_{|\psi } B[\phi, \psi] \right) - 2   m_{3/2}  {\rm Re}\left( \frac{W}{|W|} \bar{F}^{ \phi}  \nabla_{\psi } R_{\phi \bar \psi \psi \bar \psi}   \right) \nonumber \\
&+& \mgrs \Big( 6 \left(1 +  B[\phi ,\psi] \right)^2 - 2 + 3 \nabla^{2}_{|\psi } B[\phi, \psi] + B[\psi, \psi] \Big)
 \, ,
\eea
after using \eqref{Cond1} and \eqref{Cond2}. Here $ \nabla^{2}_{|\psi} B[\phi, \psi] = \tilde K^{\psi \bar \psi} \nabla^{2}_{\psi \bar \psi } B[\phi, \psi]$. Evidently, the quartic interaction term depends on the holomorphic bisectional curvature,  and thus  provide a nontrivial correlation of the scenario.

\section{\label{sec:conc} Conclusions}
The generation of the observed baryon asymmetry is one of the outstanding problems of twentieth century physics which  remains unsolved a decade into the twenty-first century. Fortunately, as cosmological observations continue to become ever more exact, one may ask to what extent the rise of precision cosmology can help  solve the question of baryogenesis.  

In this paper, we have explored a prediction of a sub-class of \ad scenarios, in which a flat direction transitions from a small vev to a larger vev during the period of inflation when cosmological scales left the horizon. In this case, we found that the near scale invariance of the cosmic microwave background places severe restrictions on the nature of the flat direction. 

A two-field system with a transitioning flat direction can be modeled ---  at least as a rough, first approximation --- as a single-field model with a step in the potential. Since the temperature spectrum of the CMB appears to favor a largely featureless inflaton potential, observations of the temperature anisotropies impose strong constraints on the location, the size and the width of any step appearing in the inflaton potential. Interestingly, the $\Lambda$CDM cosmological model with a step located at a specific position in the inflaton potential can improve the fit to WMAP data by a marginally significant amount. 

In the sub-class of models considered in \S\ref{sec:corr}, the corresponding constraints on the size and width of the step can be interpreted as constraints on the parameters of the transitioning flat direction, namely the dimension at which the flat direction is lifted in the superpotential, $n$, and the holomorphic bisectional curvature between the inflaton and the flat direction, denoted $B[\phi, \psi]$, which determines the (tachyonic) mass of the flat direction at the origin in field space.

In the toy model considered in  \S\ref{sec:num} with a Hubble parameter during inflation given by $H \simeq  10^{-6}\ \M$, \emph{no} flat direction with a tachyonic mass  in the range $-\frac{H^2}{2}$ and $-2H^2$ and $n$ in the range from $4$ to $12$ could produce a step in the inflaton potential of the size and width included within the $68\%$ confidence region of the best-fit value.   This suggests that transitioning flat directions during inflation can be severely constrained by CMB data.
However, our analysis in \S\ref{sec:num} should be regarded as a first step towards a better understanding of the cosmological predictions of this sub-class of \ad models, and a full analysis, like the one of \cite{Hunt:2004vt, Hunt:2007dn} done in the context of multiple inflation would be very interesting to pursue for a broader range of inflationary models.  
 
 Future precision cosmology observations of the E-mode spectrum and, possibly, of a non-vanishing  non-Gaussianity of the temperature anisotropies will determine the nature and significance of the features in the temperature spectrum responsible for the improved fit for a potential with a step.  Either outcome will provide important information about the naturalness of the \ad scenario. Since observations of non-Gaussianities can potentially  serve to discriminate between a single-field model with a step  and a multiple-field model with a transitioning flat direction (see e.g. \cite{Chen:2006xjb, Hotchkiss:2009pj}), it would certainly be interesting to study the non-Gaussianities produced by a transitioning field in  a variety of inflationary models.

From the point of view of string phenomenology, the \ad mechanism is a particularly attractive scenario for the generation of the baryon asymmetry. By being sensitive to Planck-suppressed operators whose structure are dictated by the string theory realization, the mechanism can potentially provide clean information about the coupling between the visible sector  and the inflationary sector. For instance, if  the \ad mechanism indeed is responsible for the observed baryon asymmetry, then it immediately follows that the early universe cannot be described by brane inflation together with a sequestered visible sector, as we have discussed in \S\ref{sec:string}.

In this paper, we have also elaborated on the possibility of extracting correlated prediction from the nontrivial structure of ${\cal N} = 1 $ supergravity. In particular we have discussed fermion couplings, multi-field couplings and higher-order Planck-suppressed interaction terms, which are completely or partially correlated with the magnitude of the tachyonic mass of the flat direction at the origin in field space through the \hbc that appear repeatedly in the supergravity Lagrangian.

An attractive feature of the  \ad mechanism is that it not only   solves the problem of baryogenesis, but that it appears to provide a robust frame-work for production of dark matter and for explaining the approximate coincidence of the dark matter and the baryon densities. If supersymmetry is relevant for the description of our universe, then  the \ad mechanism could very well  play a key role in the unification and solution of several cosmological problems.

\addtocontents{toc}{\SkipTocEntry}
\section*{Acknowledgments}
I would like to thank
   Richard Easther, Sohang Gandhi, Fawad Hassan, Bart Horn, Guy Moore, Rachel Rosen, Sara Rydbäck, Stefan Sjörs, Bo Sundborg, Philip Tanedo,   Bret Underwood and the referee  for helpful comments. 
   I am particularly grateful to Marcus Berg, Sohang Gandhi and Liam McAllister  for many insightful comments and for reading a draft of this paper.  I would also like to thank the referee for insightful comments and for several good suggestions.  Furthermore, I am  thankful to  the CoPS group at Stockholm University for their kind hospitality while this work was completed and to the organizers of the conference String Phenomenology 2011 for the opportunity to present this work. 
I gratefully acknowledge
support for this work by the Swedish Foundation for International
Cooperation in Research and Higher Education.

\vfil

\newpage
\begingroup\raggedright\endgroup

\end{document}